\documentclass[useAMS,usenatbib]{mn2e}
\usepackage{psfig}

\newcommand\apj{ApJ}%
\newcommand\apjl{ApJ}%
\newcommand\apjs{ApJS}%
\newcommand\aap{A\&A}%
\newcommand\mnras{MNRAS}%
\newcommand\pasp{PASP}%
\newcommand\nat{Nature}%
\def\subsun{\mbox{$_{\odot}$}}
\def\lesssim{\mathrel{\hbox{\rlap{\hbox{%
 \lower4pt\hbox{$\sim$}}}\hbox{$<$}}}}
\def\gtrsim{\mathrel{\hbox{\rlap{\hbox{%
 \lower4pt\hbox{$\sim$}}}\hbox{$>$}}}}

\title[Metal Cooling in Simulations of Cosmic Structure Formation]{Metal Cooling in Simulations of Cosmic Structure Formation}

\author[Britton Smith, Steinn Sigurdsson, and Tom Abel]
{Britton Smith$^{1}$\thanks{E-mail:britton@astro.psu.edu (BDS); 
steinn@astro.psu.edu (SS); tabel@stanford.edu (TA)}, 
Steinn Sigurdsson$^{1}$\footnotemark[1], and Tom Abel$^{2}$\footnotemark[1]\\
$^{1}$Astronomy \& Astrophysics, The Pennsylvania State University, 525 Davey Laboratory, 
University Park, PA 16802, U. S. A.\\
$^{2}$Kavli Institute for Particle Astrophysics and Cosmology, Stanford Linear Accelerator Center, 2575 Sand Hill Road, MS 29, \\Menlo Park, CA 94025, U. S. A.}

\begin{document}

\date{In original form May 23, 2007}

\pagerange{\pageref{firstpage}--\pageref{lastpage}} \pubyear{2007}

\maketitle

\label{firstpage}

\begin{abstract}
The addition of metals to any gas can significantly alter its evolution by 
increasing the rate of radiative cooling.  In star-forming environments, 
enhanced cooling can potentially lead to fragmentation and the formation of 
low-mass stars, where metal-free gas-clouds have been shown not to fragment.  
Adding metal cooling to numerical simulations has traditionally required a choice 
between speed and accuracy.  We introduce a method that uses the sophisticated 
chemical network of the photoionization software, Cloudy, to include radiative 
cooling from a complete set of metals up to atomic number 30 (Zn) that can be 
used with large-scale three-dimensional hydrodynamic simulations.  Our method is 
valid over an extremely large temperature range (10 K $\le$ T $\le$ 10$^{8}$ K), 
up to hydrogen number densities of 10$^{12}$ cm$^{-3}$.  At this density, a 
sphere of 1 M$\subsun$ has a radius of roughly 40 AU.  
We implement our method in the adaptive mesh refinement (AMR) 
hydrodynamic/N-body code, Enzo.  
Using cooling rates generated with this method, we study the physical conditions 
that led to the transition from Population III to Population II star formation.  
While C, O, Fe, and Si have been previously shown to make the 
strongest contribution to the cooling in low-metallicity gas, we find that up to 
40\% of the metal cooling comes from fine-structure emission by S, when solar 
abundance patterns are present.  At metallicities, Z $\ge$ 10$^{-4}$ Z$\subsun$, 
regions of density and temperature exist where gas is both thermally unstable and 
has a cooling time less than its dynamical time.  We identify these doubly 
unstable regions as the most inducive to fragmentation.  At high redshifts, the 
cosmic microwave background inhibits efficient cooling at low temperatures and, 
thus, reduces the size of the doubly unstable regions, making fragmentation 
more difficult.
\end{abstract}


\section{Introduction}

The first luminous objects in the universe formed from primordial gas, comprised 
solely of H and He, with only trace amounts of D an Li.  The relatively simple 
chemistry of metal-free gas, combined with tighly constrained cosmological 
parameters \citep{2007ApJS..170..377S}, has allowed the formation of the first 
stars to be simulated with extremely high precision, from the hierarchical growth 
of their host dark matter halos through to the point where the dense proto-stellar 
cores becomes optically thick \citep{2002Sci...295...93A,2002ApJ...564...23B,
2004NewA....9..353B,2006ApJ...652....6Y,2007ApJ...654...66O,2007MNRAS.378..449G}.  
With the deaths of these stars came the creation of the first heavy elements.  
Core-collapse and pair-instability supernovae created metals in copious amounts 
\citep{2002ApJ...567..532H} and ejected them into the IGM 
\citep{2001ApJ...555...92M}.

The presence of metals alters the dynamics of collapsing gas-clouds by increasing 
the number of available atomic and molecular transitions, allowing the gas to 
lose its internal energy more quickly than in case of no metals 
\citep{2000ApJ...534..809O,2001MNRAS.328..969B,2003Natur.425..812B}.  The 
introduction of metals adds a new level of complexity to the problem of 
simulating the formation and evolution of cosmic structure.  
\citet{1997NewA....2..181A} identified a minimal set of 21 chemical reactions 
necessary for accurately following the non-equilibrium evolution of a gas consisting 
solely of species of H and He, including H$_{2}$.  \citet{1998A&A...335..403G} 
showed that 33 total reactions were required when including D and Li species to the 
gas.  \citet{2000ApJ...534..809O} performed one of the first numerical studies of 
collapsing gas-clouds to consider the contribution of metals.  Their chemical 
network of H, He, C, and O included 50 atomic and molecular species and 478 reactions.  
While theirs was not a minimal 
model, the above examples illustrate the great expense associated with the expansion 
of chemical networks to include additional elements.  Other works have studied the 
effect of metals on star-forming gas using similar methodologies to that of 
\citet{2000ApJ...534..809O}, e.g., \citet{2002ApJ...571...30S,2003Natur.422..869S,
2006MNRAS.369.1437S,2005ApJ...626..627O}.  The complexity of the chemical networks 
used in these studies limited their treatment of gas evolution to one-zone, 
semi-analytical models.  In the earliest work to incorporate metal cooling into 
three-dimensional hydrodynamic simulations to study metal-enriched star formation, 
\citet{2001MNRAS.328..969B} used a small set of the most dominant atomic 
transitions of C, N, O, Fe, Si, and S, as decribed by \citet{1997ApJ...485..254R}.  
Their method also ignored the cooling from H$_{2}$, which was justified within 
their study by the assumption of a very large photo-dissociating UV background, 
but is 
otherwise an extremely important coolant in low-metallicity environments.  For 
high temperature gases, \citet{1993ApJS...88..253S} computed metal cooling 
functions that included 14 heavy elements over a range of metallicities, 
with solar abundance patterns.  These cooling functions are useful for simulating 
the IGM and other hot, ionized environments, but a minimum temperature of 
10$^{4}$ K makes them inapplicable to studies of the cold, neutral gas associated 
with star-formation.  These cooling functions assume collisional equilibrium of 
the species and as such cannot capture the important role of UV and X-ray radiation.

We introduce a new method for including the cooling from heavy elements in 
large-scale hydrodynamic simulations that is valid over a wide range of physical 
conditions, covers a great number of elemental species, and is fast enough to be 
used in large-scale numerical simulations.  We have utilized the established 
photoionization software, Cloudy \citep{1998PASP..110..761F} to construct large 
grids of metal cooling data.  We have developed a method to include both the 
cooling from heavy elements and the non-equilibrium cooling from H$_{2}$ in 
hydrodynamic simulations.  This method has been used successfully in the numerical 
simulations of star formation performed by \citet{2007ApJ...661L...5S}.  
In \S\ref{sec:method}, we describe our method for 
creating the metal cooling data, including a new code to expedite the process.  
We, then, present two implementations of the cooling method in the AMR, 
hydrodynamic/N-body code, Enzo \citep{1997WSAMRGMBryan,2004CWAMROShea}.  
In \S\ref{sec:metals}, we focus on the application of metals to low-temperature 
environments, identifying the dominant cooling mechanisms, and studying the 
possibility of fragmentation and thermal instability in metal-enriched gas.  
Finally, we end with a discussion in \S\ref{sec:discussion} of the role played by 
the heavy elements in the formation of structure in the early universe.

\section{Numerical Method} \label{sec:method}

\subsection{Calculation of Metal Cooling Rates} \label{sec:cloudy}

At the current time, it is still too computationally expensive and 
memory intensive to follow the 
non-equilibrium chemistry for a large set of heavy elements in a three-dimensional 
hydrodynamic simulation.  
The exact mass of the first massive stars is not known 
\citep{2002Sci...295...93A,2004ApJ...603..383T,2006ApJ...652....6Y}.  
Also unknown are the exact yields of early supernovae 
\citep{2002ApJ...567..532H,2005ASPC..336...79M,2006NuPhA.777..424N,
2006astro.ph..8028R}.  Similarly, in many astrophysical systems one might want to 
model computationally the exact metal distributions. Consequently, it is not clear 
a priori what level of sophistication of cooling model is needed to adequately 
capture the hydro and thermodynamic evolution of the gas under consideration.  
Note that uncertain grain physics also increases the potentially  
important parameter space.  
In our approach, we assume ionization equilibrium, which allows us to calculate, 
in advance, the cooling rate for a parcel of gas with a given density and 
temperature, with incident radiation of known spectral shape and intensity.  
For this problem, we find the photoionization code, Cloudy 
\citep{1998PASP..110..761F}, especially apt.  
Cloudy is conventionally used to model the transmitted spectrum from a cloud of 
gas with a given chemical composition, being irradiated by a specified source.  
The code must calculate an equilibrium solution by balancing the incident heating 
with the radiative cooling from a full complement of atomic and molecular 
transitions, as well as continuum emission from dust.  
The chemical network of Cloudy covers all atomic species from H to Zn, as well as 
a multitude of molecular species.  
Each elemental abundance can be specified individually, giving us the ability 
to model the cooling from a gas with any composition.  
Since Cloudy permits the use of virtually any input spectrum, we are able to create 
cooling data that is suitable for any radiation environment.  
Instead of allowing the code to cycle through temperatures until converging on a 
thermodynamic equilibrium solution, we 
use the \texttt{constant temperature} command to fix the temperature externally, 
allowing us to utilize Cloudy's sophisticated machinery to calculate 
cooling rates out of thermal equilibrium.
In this manner, we create a grid of heating and cooling values as a function of 
temperature, gas density, chemical composition, and incident spectrum.  The cooling 
rates presented in this work were created using version 07.02.01 of the Cloudy 
software.

To automate the process of data production and organization, we have created a 
code, called ROCO (Recursively Organized Cloudy Output.)  ROCO uses a recursive 
algorithm to process user-specified loop parameters, making it possible to 
create data-grids of any dimension.  
Commands that are to be issued to Cloudy are given to the ROCO code in either one 
of two formats - loop commands with a set of parameters through which the code will 
iterate, and constant commands that are to be issued with the same value during 
each iteration over the loop commands.  
Since most uses of Cloudy involve the creation of large grids of models constructed 
by looping over a set of input parameters, the capabilities of ROCO give it the 
potential to be useful to a broader community of Cloudy users than just those who 
would use it to create the cooling tables discussed here.  
To this end, ROCO is structured in such a way that the post-Cloudy data analysis 
routines can be easily interchanged to suit the needs of different users.  
The code features an extra running mode that simply runs Cloudy over the specified 
parameter-space with no further processing of the data, as well as a template 
designed to help users create new running modes suited to their specific needs.  
ROCO also has the ability to run multiple instances of Cloudy simultaneously, 
greatly reducing runtime.  
The parallel feature works well on individual machines with multiple processors, as 
well as Beowulf clusters using the MPI (Message Passing Interface) framework.  
A copy of the ROCO code will be made available upon request to the authors.

\begin{figure}
  \psfig{figure=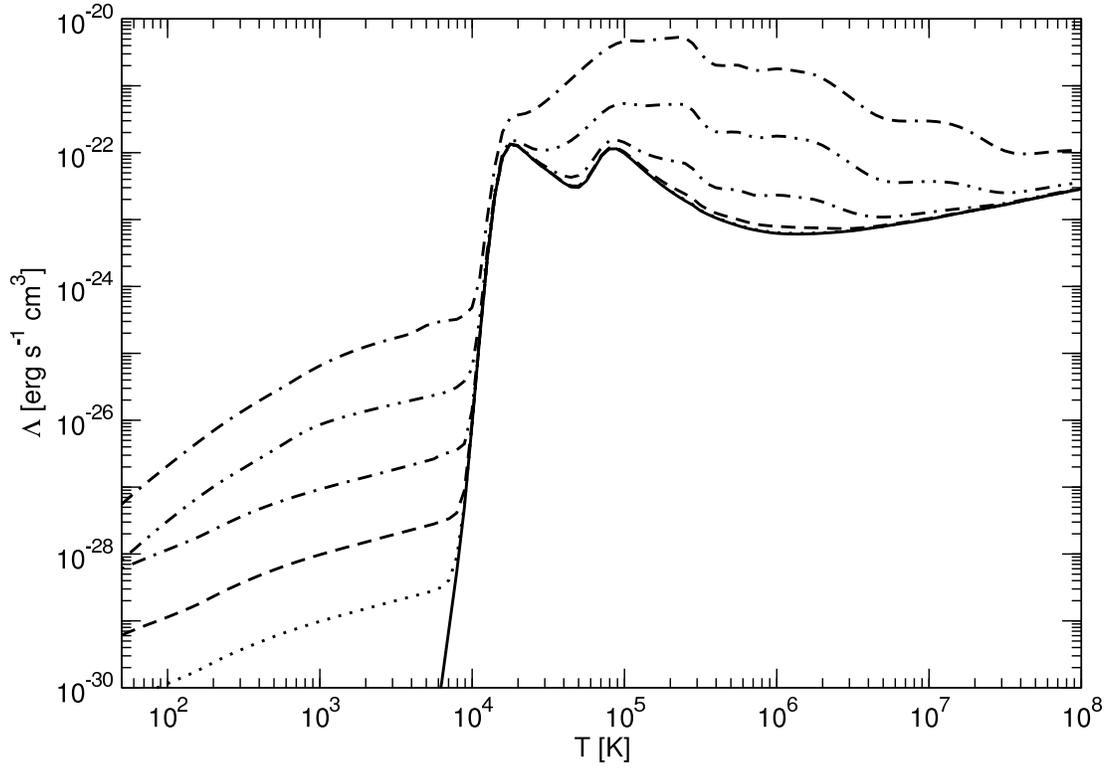,angle=-90,width=6.5in}
  \caption{Cooling functions, excluding cooling from H$_{2}$, for gases 
    with n$_{H}$ = 1 cm$^{-3}$ and metallicities, Z = 0 (solid), 10$^{-3}$ 
    Z$\subsun$ (dot), 10$^{-2}$ Z$\subsun$ (dash), 10$^{-1}$ Z$\subsun$ 
    (dot-dash), 1 Z$\subsun$ (dot-dot-dash), and 10 Z$\subsun$ (dot-dash-dash).
  } \label{fig:cooling1}
\end{figure}

In Figure \ref{fig:cooling1}, we display the resulting cooling function for gas 
with n$_{H}$ = 1 cm$^{-3}$ at metallicities, from Z = 0 (metal-free) to 10 
Z$\subsun$, over the temperature range, 50 $\le$ T $\le$ 10$^{8}$ K.  For these 
cooling rates, we use the \texttt{coronal equilibrium} 
command in Cloudy to simulate an environment free of radiation, where all 
ionization is collisional.  We also neglect the cooling from H$_{2}$, so as to 
better illustrate the cooling contribution from metals at temperatures less than 
10$^{4}$ K.  We accomplish this by issuing the Cloudy command, 
\texttt{no H2 molecule}.

\subsection{Implementation in Hydrodynamic Simulations} \label{sec:enzo}

We implement our metal cooling method in the Eulerian adaptive mesh refinement 
hydrodynamic/N-body code, Enzo\citep{1997WSAMRGMBryan,2004CWAMROShea}.  
When a simulation is initialized, Enzo reads in the Cloudy/ROCO data-grid, storing 
the heating and cooling values as functions of temperature, H number density, and 
any other parameters, such as spectral intensity, depending on the nature of the 
simulation.  
The heating, $\Gamma$, and cooling, $\Lambda$, are stored with code units 
corresponding to [ergs s$^{-1}$ cm$^{3}$].  
During the simulation, Enzo stores the mass density and internal energy for each 
grid cell in the box.  
At each hydrodynamic time-step, the radiative cooling solver cools the gas by 
lowering the internal energy via a simple Euler update, 
\begin{equation} \label{eqn:cool}
u_{i,j,k}^{n+1} = u_{i,j,k}^{n} + \dot{u}_{i,j,k}^{n} \times \delta t,
\end{equation}
where $u_{i,j,k}^{n+1}$ denotes the internal energy of the grid cell with (x,y,z) 
coordinates, (i,j,k), at the (n+1)'th time-step, $\dot{u}$ is the cooling 
rate in code units corresponding to [ergs s$^{-1}$], and $\delta t$ is the adopted 
time-step.  For every hydrodynamic time-step, the code subcycles 
through Equation \ref{eqn:cool}, selecting from three possible time-steps, 
until one full hydrodynamic time-step has been completed.  The time-step, $\delta t$, 
in Equation \ref{eqn:cool}, adopts the minimum of the following three values: 
(1) half of the hydrodynamic time-step, (2) 10\% of the cooling time, (u/$\dot{u}$), 
or (3) the time remaining to have integrated over one full hydrodynamic time-step.  
The Euler update is the 
standard method used for updating the internal energy in all of the established 
cooling routines in the Enzo code.  Since the cooling rate is such a nonmonotonic 
function of temperature, this approach of using substeps with an explicit integration 
method has been found to yield the best combination of accuracy and speed 
\citep{1994ApJ...429..434A}.  
The internal energy and mass density for each grid cell are converted to 
temperature and number density and the heating and cooling values are calculated 
by linearly interpolating over the Cloudy/ROCO data-grid.  
The change in internal energy from the Cloudy/ROCO cooling rates is expressed as
\begin{equation} \label{eqn:dudt}
\dot{u}_{C/R} = (\Gamma - \Lambda) n_{H},
\end{equation}
where n$_{H}$ is the H number density.  

We implement two distinct versions of the method described above.  
In the first and simplest version, the cooling is calculated solely from the 
Cloudy/ROCO data, as in Equation \ref{eqn:dudt}.  The total change in internal 
energy is
\begin{equation} \label{eqn:version1}
\dot{u}_{tot} = \dot{u}_{C/R}.
\end{equation}
When converting the internal energy to temperature, it is necessary to know the 
value of the mean molecular weight, $\mu$.  In this implementation, we assume $\mu$ 
to be a constant with the value 1.22.  
For high temperatures, T $\gtrsim$ 10$^{4}$ K, this method is sufficient for 
providing accurate gas cooling within hydrodynamic simulations.  This 
implementation is not suitable, however, when T $\lesssim$ 10$^{4}$ K and the 
formation of H$_{2}$ becomes important.  Disregarding formation on grain surfaces 
and three-body formation, H$_{2}$ primarily forms through 
the following channels: \\
the H$^{-}$ channel,
\begin{equation} \label{eqn:hminus}
\begin{array}{ll}
\textrm{H + e$^{-}$ $\rightarrow$ H$^{-}$ + $\gamma$,}\\
\textrm{H$^{-}$ + H $\rightarrow$ H$_{2}$ + e$^{-}$,}\\
\end{array}
\end{equation}
and the H$_{2}^{+}$ channel,
\begin{equation} \label{eqn:h2plus}
\begin{array}{ll}
\textrm{H + H$^{+}$ $\rightarrow$ H$_{2}^{+}$ + $\gamma$,}\\
\textrm{H$_{2}^{+}$ + H $\rightarrow$ H$_{2}$ + H$^{+}$.}\\
\end{array}
\end{equation}
When a significant electron fraction exists, these reactions proceed to form 
H$_{2}$ very quickly, with the H$^{-}$ channel typically dominating, except in 
the very high redshift universe (z $>$ 100), where H$^{-}$ is readily destroyed by 
the CMB \citep{1997NewA....2..181A,2002ApJ...564...23B}.  
Recently, \citet{2006MNRAS.372.1175H} have suggested that formation of H$_{2}^{+}$ 
via
\begin{equation} \label{eqn:hehplus}
\begin{array}{ll}
\textrm{He + H$^{+}$ $\rightarrow$ HeH$^{+}$ + $\gamma$,}\\
\textrm{HeH$^{+}$ + H $\rightarrow$ H$_{2}^{+}$ + He,}\\
\end{array}
\end{equation}
is responsible for more H$_{2}$ than the H$_{2}^{+}$ channel in Equation 
\ref{eqn:h2plus}.  
If, however, the ionization fraction is low, H$_{2}$ forms very slowly, with 
equilibrium timescales that can exceed the current age of the universe.  
The consequence is that H$_{2}$ formation is so sensitive to the thermal history 
of the gas that H$_{2}$ fractions cannot be known without explicitly following the 
non-equilibrium chemistry during the simulation.  
We find this to be the case when using Cloudy to compute the cooling rate from 
H$_{2}$.  
In searching for ionization equilibrium, Cloudy integrates over timescales that are 
unphysically long, leading to an overcalculation of the H$_{2}$ fraction, 
producing cooling rates that are too high.  

We solve this problem in our second implementation by first removing the H$_{2}$ 
molecule from Cloudy's chemical network with the \texttt{no H2 molecule} command.  
We, then, run two Cloudy/ROCO data-grids: one with the full set of elements and the 
other with H and He only.  We obtain a metals-only data-grid by subtracting the 
metal-free data-grid from the complete data-grid.  Using the established H/He 
network in Enzo \citep{1997NewA....2..209A,1997NewA....2..181A}, we follow the 
non-quilibrium fractions of H, H$^{+}$, H$^{-}$, H$_{2}$, H$_{2}^{+}$, He, He$^{+}$, 
He$^{++}$, and e$^{-}$, and directly calculate the associated atomic 
\citep{1981MNRAS.197..553B,1992ApJS...78..341C} and molecular 
\citep{1998A&A...335..403G} cooling rates.  We 
provide the option to use the H$_{2}$ cooling rates of \citet{1983ApJ...270..578L}, 
which are obsolete, but allow a means of comparison to older simulations.  
We also include the cooling, or heating, from electrons scattering off the CMB as
\begin{equation} \label{eqn:comp}
\Lambda_{Comp} = 5.4 \times 10^{-36}(1 + z)^{4}n_{e}(T - T_{CMB}),
\end{equation}
where T$_{CMB}$ = 2.7 (1 + z) K \citep{2002ApJ...564...23B}.
We prevent the metals from cooling the gas below the CMB temperature by 
subtracting the metal cooling rate at T = T$_{CMB}$, as in 
\citet{2001MNRAS.328..969B}.  
Including the CMB explicitly 
in cosmological simulations would require adding an extra dimension in the 
Cloudy/ROCO data-grid to account for the evolution of the CMB with redshift.  
While this is certainly possible, interpolating over an extra dimension to 
calculate the cooling during the simulation would be significantly slower than 
the approximation described above.  
To test the validity of this approximation of the CMB temperature-floor, we 
create a Cloudy/ROCO metals-only data-grid, explicitly including the CMB.  
At low densities (n $\sim$ 1 cm$^{-3}$), 
the values of ($\Lambda$ - $\Gamma$) from the data with the CMB included 
differ from the values of ($\Lambda$ - $\Lambda(T_{CMB})$) from the data 
without the CMB by a factor of roughly 2 near T$_{CMB}$.  At higher temperatures 
and densities, the two values are nearly identical.  
The total rate of energy loss applied to the simulation gas in the second 
implementation is 
\begin{equation} \label{eqn:version2}
\dot{u}_{tot} = \dot{u}_{H,He} + \dot{u}_{Comp} + \dot{u}_{C/R},
\end{equation}
where $\dot{u}_{H,He}$ is the total atomic and molecular cooling from the H/He 
network, and $\dot{u}_{C/R}$ is the metals-only cooling rate taken from the 
Cloudy/ROCO data in the manner described above.  
The value of $\mu$ is calculated directly from the H/He species fractions, 
neglecting any addition from the metals.  In low-metallicity gases, this approach is 
reasonable, as the increase in $\mu$ from the metals only reaches $\sim$10$^{-4}$ 
for Z = 10$^{-2}$ Z$\subsun$.  Since the Cloudy/ROCO data-grids also store the 
values of $\mu$ for each point, this can be added to the value calculated without 
the heavy elements when the metallicity is very high.  

\begin{figure}
  \psfig{figure=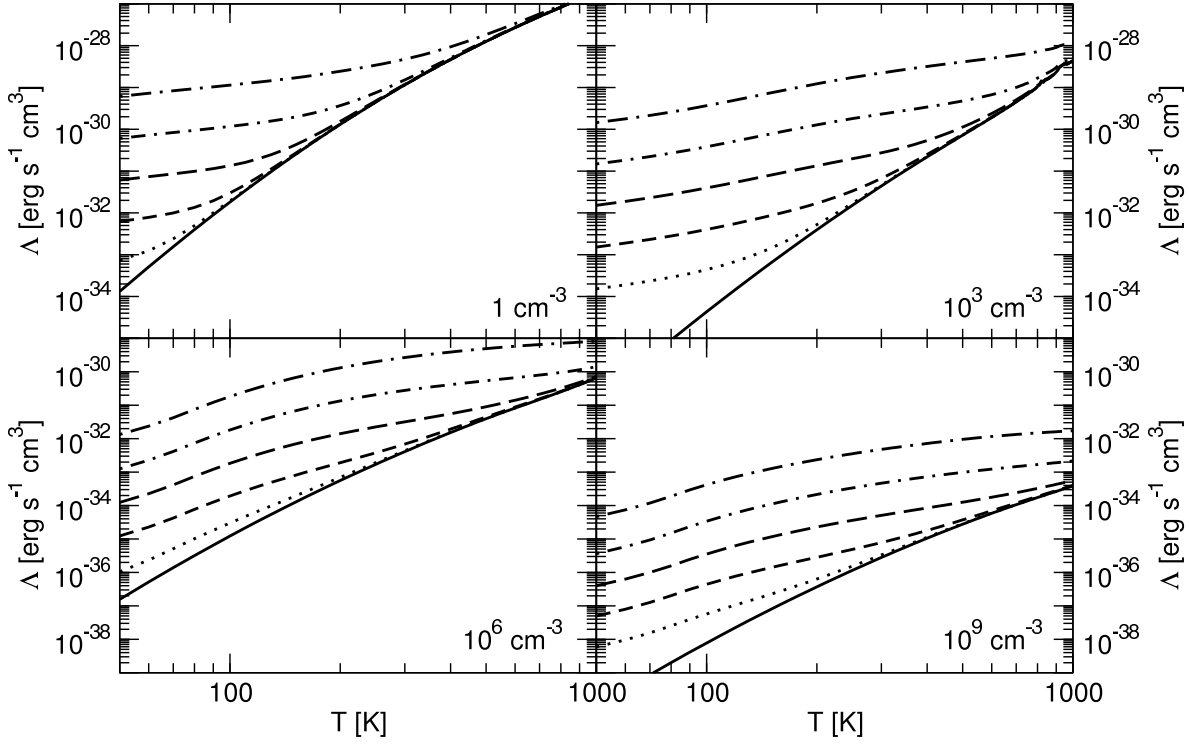,angle=-90,width=6.5in}
  \caption{Cooling functions, including H$_{2}$ cooling, for gases with 
    n$_{H}$ = 1 cm$^{-3}$ (top-left), n$_{H}$ = 10$^{3}$ cm$^{-3}$ (top-right), 
    n$_{H}$ = 10$^{6}$ cm$^{-3}$ (bottom-left), and n$_{H}$ = 10$^{9}$ 
    cm$^{-3}$ (bottom-right).  Metallicities are Z = 0 (solid), 
    10$^{-6}$ Z$\subsun$ (dot), 10$^{-5}$ Z$\subsun$ (dash), 
    10$^{-4}$ Z$\subsun$ (long dash), 10$^{-3}$ Z$\subsun$ (dot-dash), and 
    10$^{-2}$ Z$\subsun$ (dot-long dash).
  } \label{fig:cooling2}
\end{figure}

\begin{figure}
  \psfig{figure=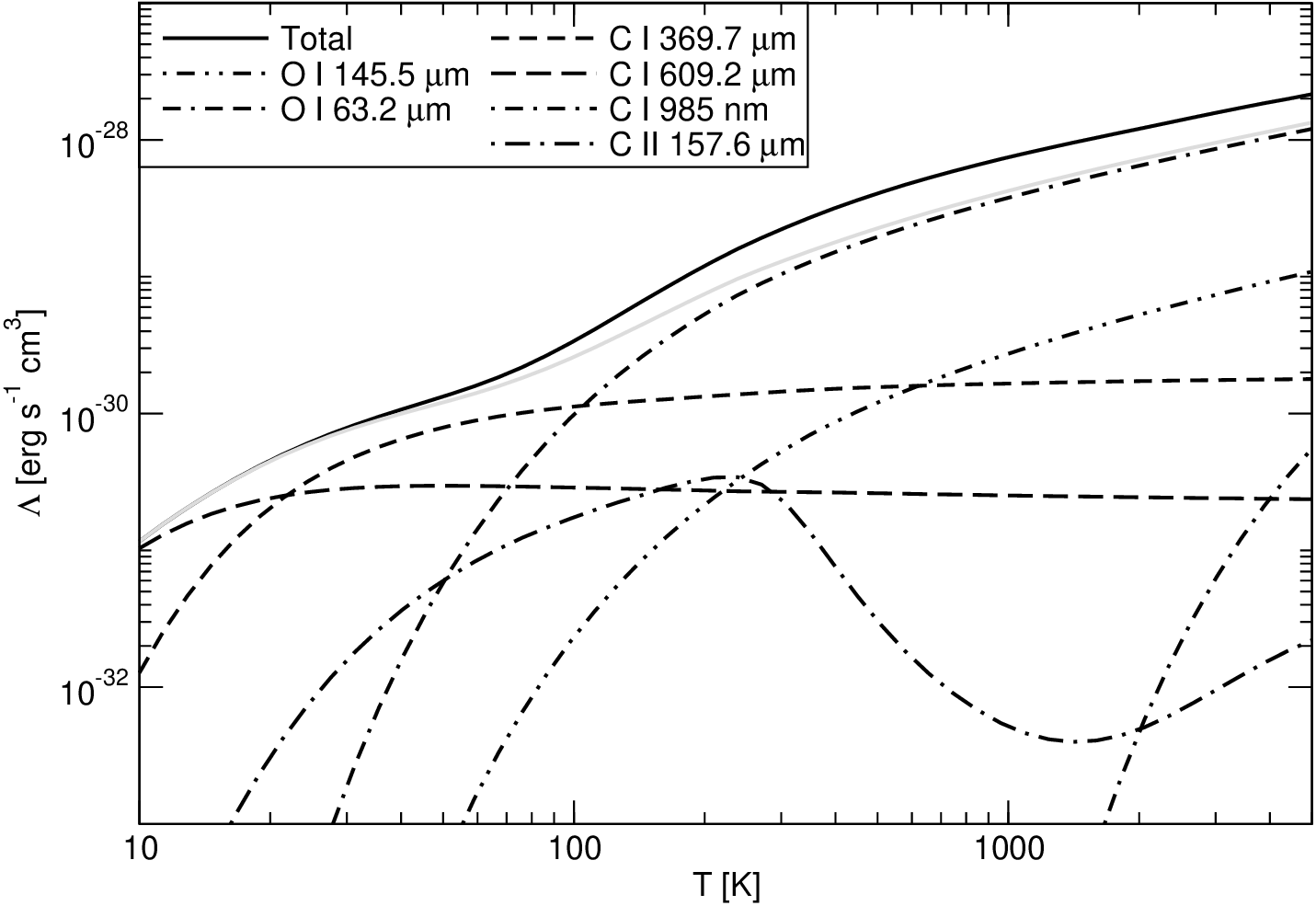,width=6.5in}
  \caption{Cooling contributions from C and O species that reach at least 
    10$^{-3}$ of the total cooling for gas with with n$_{H}$ = 10$^{3}$ cm$^{-3}$ 
    and Z = 10$^{-2}$ Z$\subsun$.  
    The total cooling (solid, black) includes all species contained within the 
    Cloudy chemical network.  
    Components shown are 
    [O\textsc{I}] 145.5 $\mu$m (dot-dot-dash), 
    [O\textsc{I}] 63.2 $\mu$m (dash-dash-dot), 
    [C\textsc{I}] 369.7 $\mu$m (dash), 
    [C\textsc{I}] 609.2 $\mu$m (long dash), 
    C\textsc{I} 985 nm (dash-dot), and 
    [C\textsc{II}] 157.6 $\mu$m (long dash-dot).  The solid, grey line indicates 
    the sum of all the components plotted.
  } \label{fig:COcooln3_Z-2}
\end{figure}

\begin{figure}
  \psfig{figure=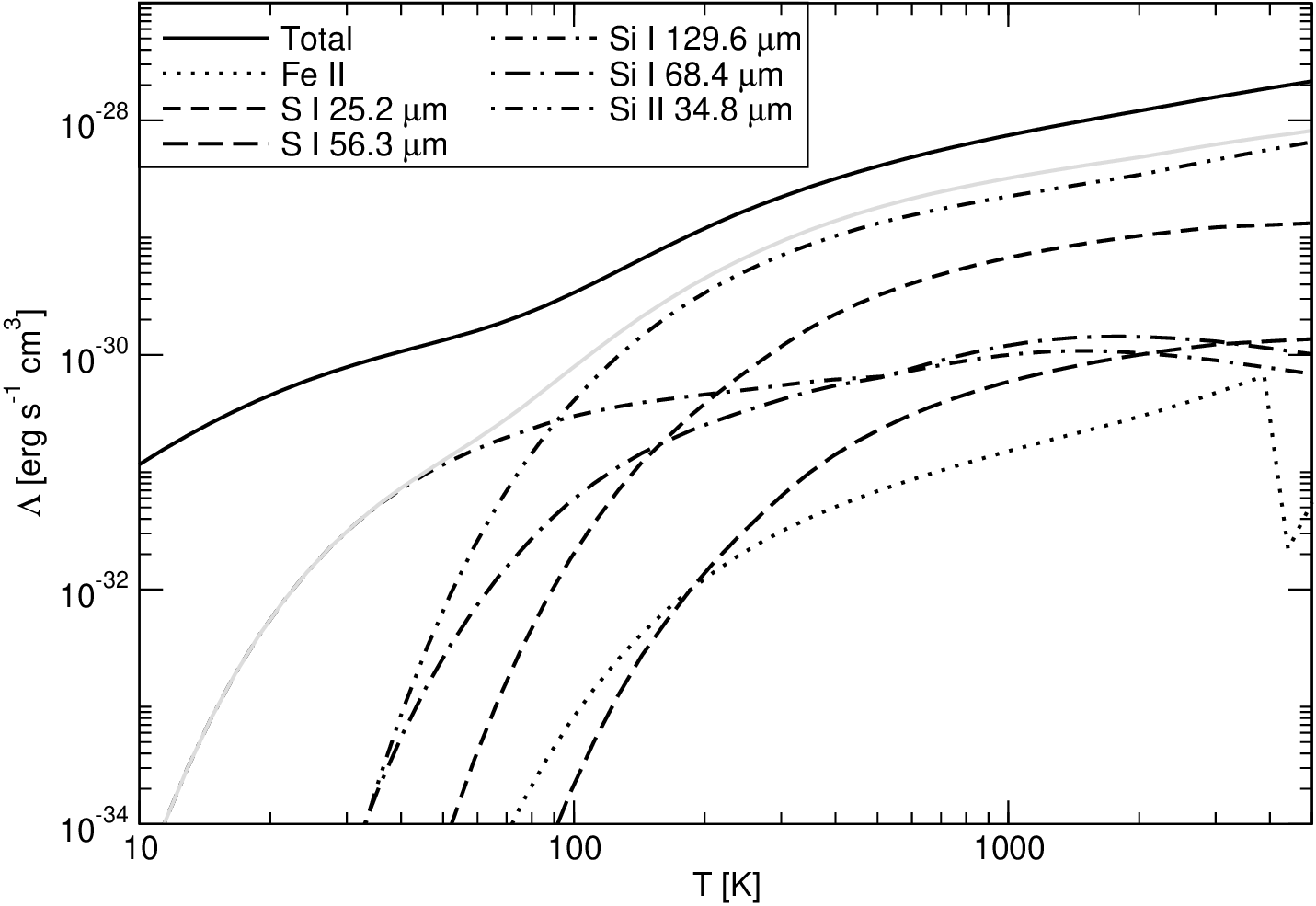,width=6.5in}
  \caption{All other coolants not plotted in Fig. \ref{fig:COcooln3_Z-2} that reach 
    at least 10$^{-3}$ of the total cooling for gas with n$_{H}$ = 
    10$^{2}$ cm$^{-3}$ and Z = 10$^{-2}$ Z$\subsun$.  The total cooling (solid, 
    black) includes all species contained within the Cloudy chemical network.  
    Components shown are [Fe\textsc{II}] (dot), 
    [S\textsc{I}] 25.2 $\mu$m (dash), 
    [S\textsc{I}] 56.3 $\mu$m (long dash), 
    [Si\textsc{I}] 129.6 $\mu$m (dash-dot), 
    [Si\textsc{I}] 68.4 $\mu$m (long dash-dot), and 
    [Si\textsc{II}] 34.8 $\mu$m (dash-dot-dot).  The solid, grey line indicates 
    the sum of all the components plotted.
  } \label{fig:FeSSicooln3_Z-2}
\end{figure}

In Figure \ref{fig:cooling2}, we display low-temperature cooling functions for 
gases with varying density and metallicity, constructed with the third 
implementation of the metal cooling method.  
To produce the data for Figure \ref{fig:cooling2}, we set up an unphysical, 
two-dimensional grid in Enzo that varies smoothly over density and temperature.
We iterate the reaction network for a time equivalent to that between 
z = 99 and 20 ($\sim$160 Myr), with hydrodynamics disabled, then compute the 
cooling with the third implementation of our metal cooling method, using the 
H$_{2}$ cooling rates of \citet{1998A&A...335..403G}.  
Since the first stars are predicted to form at the centers of $\sim$10$^{6}$ 
M$\subsun$ dark matter halos at z $\sim$ 20 \citep{1997ApJ...474....1T,
2003ApJ...592..645Y}, 
integrating the rate equations over this time interval places each of the species 
in the relative abundances in which they would be found during the epoch of 
first-star formation.  Hence, Figure \ref{fig:cooling2} provides a direct 
comparison of the cooling rate of the gas that formed the first and successive 
generations of stars.  The H$_{2}$ fractions used to create the cooling rates for Figure 
\ref{fig:cooling2} result from integrating the H/He chemical network for the period of time
between z = 99 and 20.  As such, Figure \ref{fig:cooling2} should not be used as a general 
cooling function, except in the context mentioned above.  However, Figures 
\ref{fig:COcooln3_Z-2}--\ref{fig:cool_dust} do not include the cooling from H$_{2}$ and may, therefore, 
be used as general purpose cooling functions when the cooling from H$_{2}$ is not needed.

\section{Metals in Low-Temperature Gases} \label{sec:metals}

\subsection{Dominant Coolants} \label{sec:coolants}

Much attention has been given recently to the role of the first heavy elements 
in transitioning from the singular, high-mass mode of star formation of the first 
stars to the mode producing stars with a Salpeter initial mass function (IMF).  
Analytical studies by \citet{2003Natur.425..812B} and \citet{2006ApJ...643...26S} 
have focused on the contributions of individual elements toward triggering 
fragmentation in star-forming clouds.  \citet{2003Natur.425..812B} suggest C and 
O to be the dominant coolants in low-metallicity gas, in the presence of an H$_{2}$ 
dissociating UV background created by the first stars \citep{2003ApJ...596...34B}.  
By calculating the cooling rate necessary to equate the cooling time to the 
free-fall time at n = 10$^{4}$ cm$^{-3}$ and T = 200 K, the point where H$_{2}$ 
cooling becomes inefficient \citep{2002Sci...295...93A,2002ApJ...564...23B}, 
\citet{2003Natur.425..812B} predict individual critical abundances of C and O 
to be [C/H]$_{crit}$ $\simeq$ -3.5 and [O/H]$_{crit}$ $\simeq$ -3.05, where 
[A/H] = log$_{10}$(N$_{A}$/N$_{H}$) - log$_{10}$(N$_{A}$/N$_{H}$)$\subsun$.  
\citet{2006ApJ...643...26S} consider the cooling from Fe and Si, in addition 
to C and O, and take into account the density dependence of metal cooling.  
In doing so, they find that the critical abundance of each element varies with 
density, reaching a minimum at a critical density that is different in each case.  
They also note that different elements dominate different density and 
temperature regimes.

\begin{figure}
  \psfig{figure=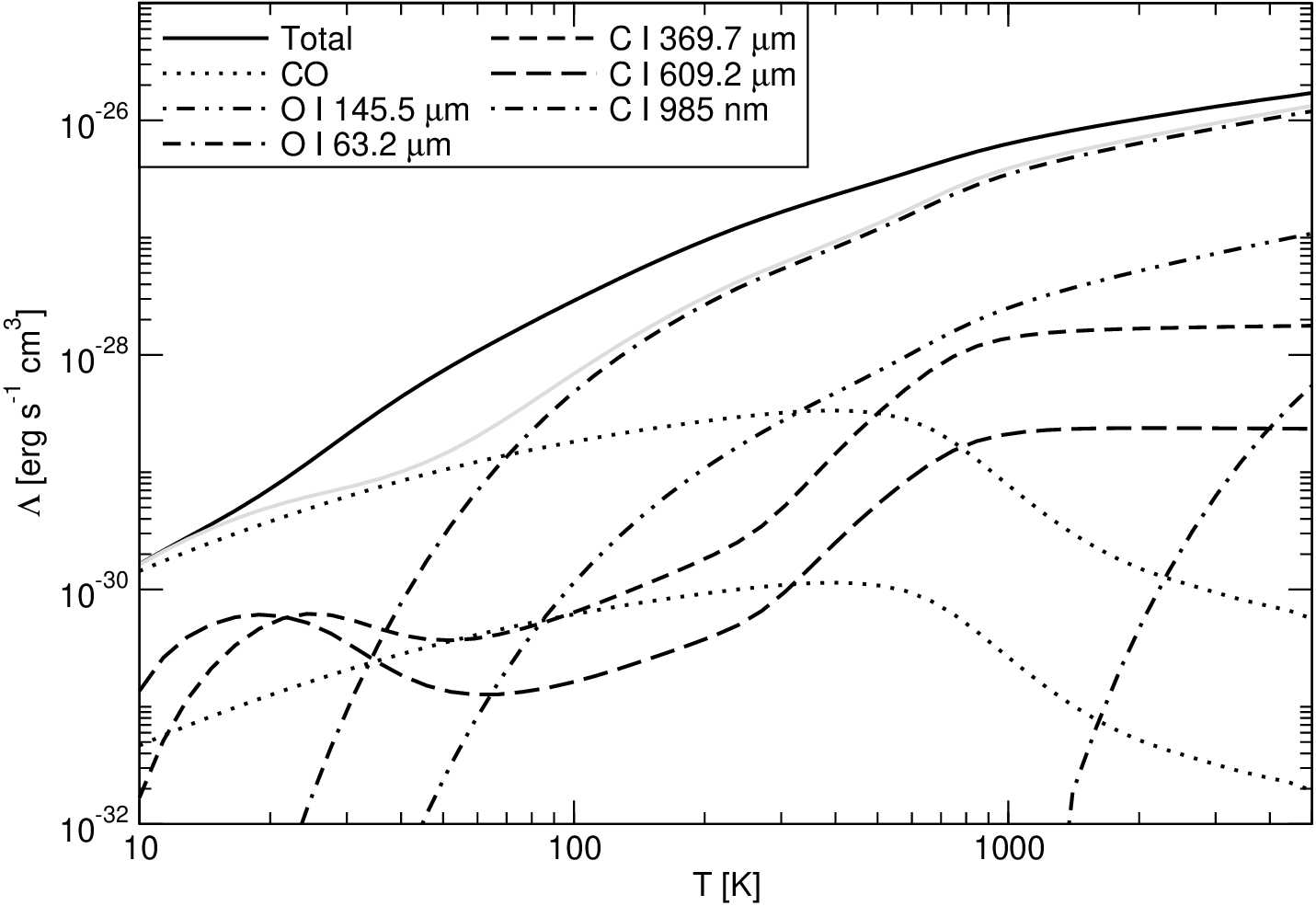,width=6.5in}
  \caption{Cooling contributions from C and O species that reach at least 
    10$^{-3}$ of the total cooling for gas with with n$_{H}$ = 10$^{3}$ cm$^{-3}$ 
    and Z = 1 Z$\subsun$.  
    The total cooling (solid, black) includes all species contained within the 
    Cloudy chemical network.  
    Components shown are CO (dot),
    [O\textsc{I}] 145.5 $\mu$m (dot-dot-dash), 
    [O\textsc{I}] 63.2 $\mu$m (dash-dash-dot), 
    [C\textsc{I}] 369.7 $\mu$m (dash), 
    [C\textsc{I}] 609.2 $\mu$m (long dash), and 
    C\textsc{I} 985 nm (dash-dot).  
    The higher dotted line represents CO emission from C$^{12}$O$^{16}$, while the 
    lower dotted line shows emission from C$^{13}$O$^{16}$.  
    The solid, grey line indicates the sum of all the components plotted.
  } \label{fig:COcooln3_Z0}
\end{figure}

\begin{figure}
  \psfig{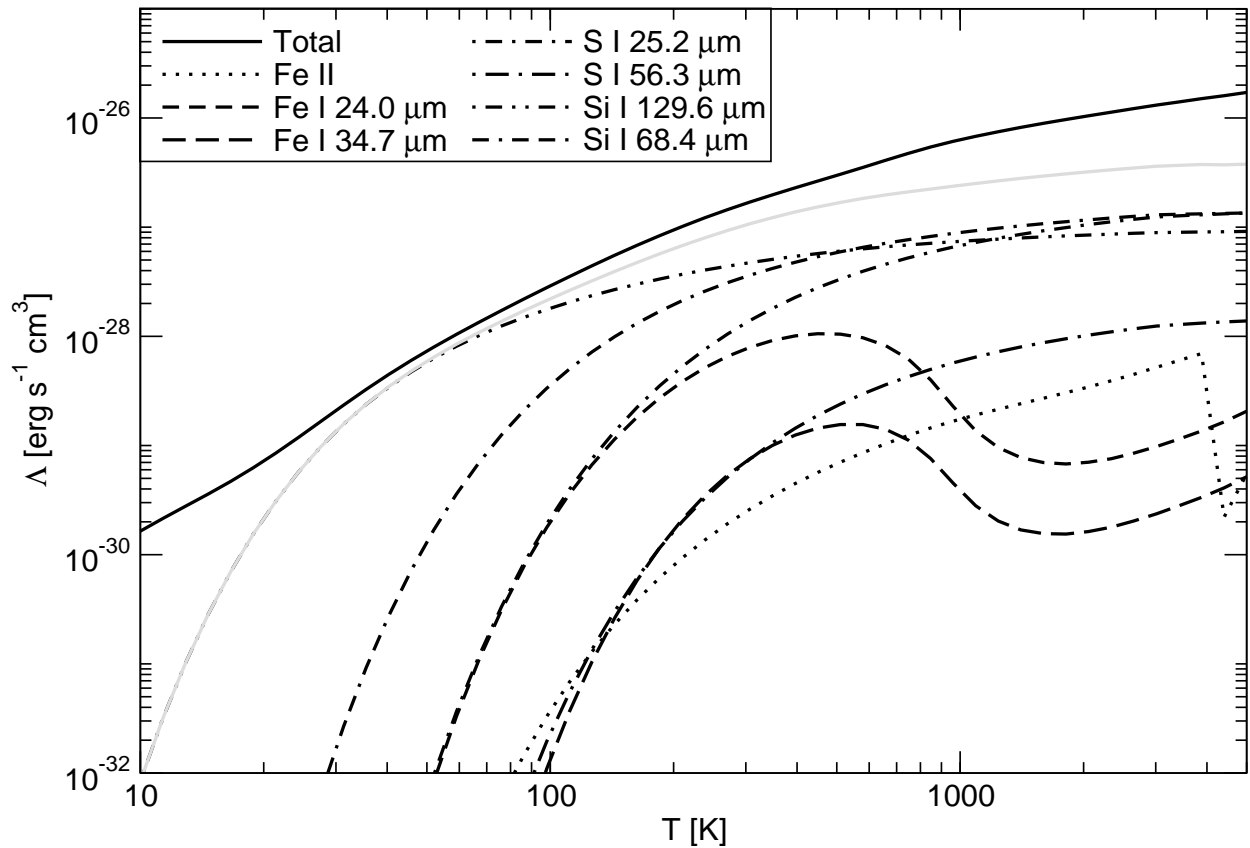}
  \caption{All other coolants not plotted in Fig. \ref{fig:COcooln3_Z0} that reach 
    at least 10$^{-3}$ of the total cooling for gas with n$_{H}$ = 
    10$^{2}$ cm$^{-3}$ and Z = 1 Z$\subsun$.  The total cooling (solid, 
    black) includes all species contained within the Cloudy chemical network.  
    Components shown are [Fe\textsc{II}] (dot), 
    [Fe\textsc{I}] 24.0 $\mu$m (dash), 
    [Fe\textsc{I}] 34.7 $\mu$m (long dash), 
    [S\textsc{I}] 25.2 $\mu$m (dash-dot), 
    [S\textsc{I}] 56.3 $\mu$m (long dash-dot), 
    [Si\textsc{I}] 129.6 $\mu$m (dash-dot-dot), and 
    [Si\textsc{I}] 68.4 $\mu$m (dash-dash-dot).  
    The solid, grey line indicates the sum of all the components plotted.
  } \label{fig:FeSSicooln3_Z0}
\end{figure}

\begin{figure}
  \psfig{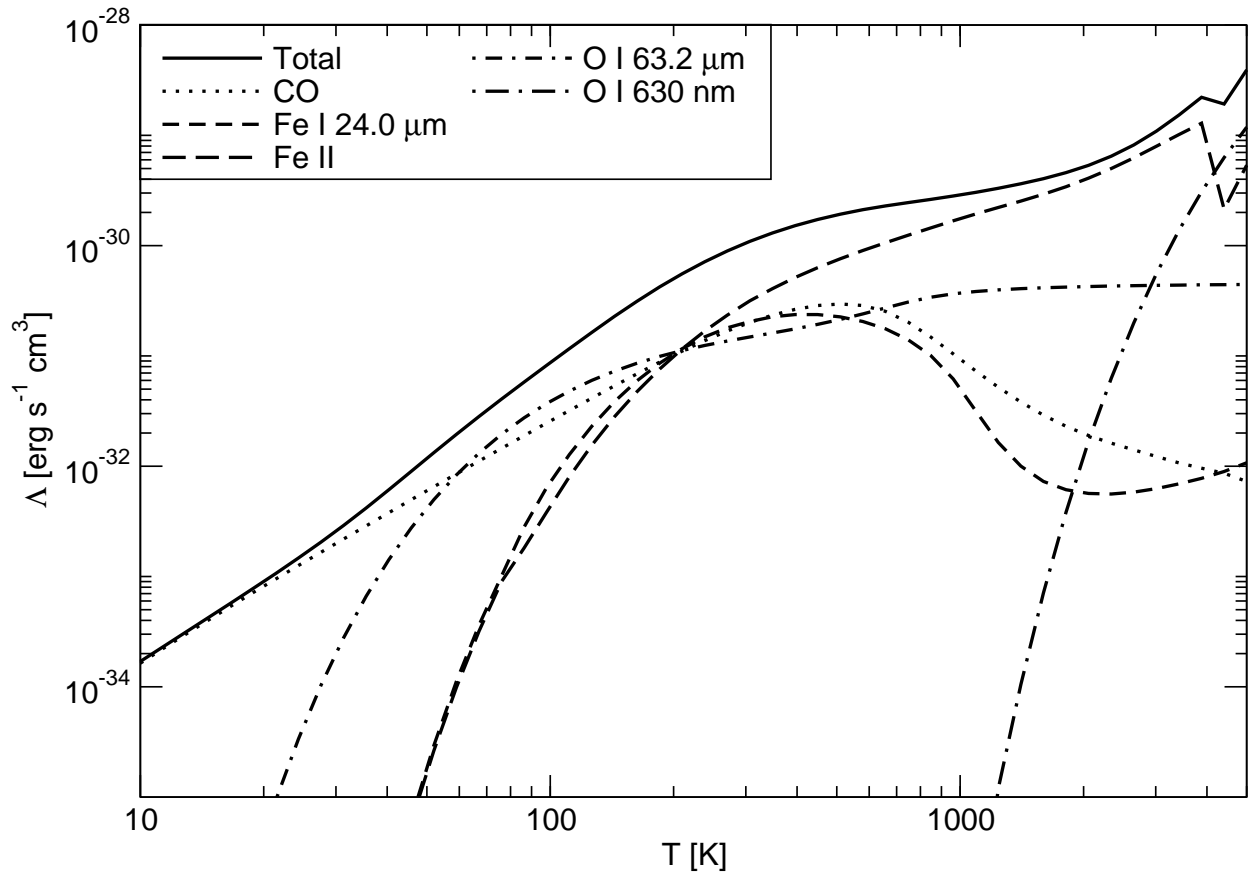}
  \caption{Subset of the most dominant coolants at n$_{H}$ = 10$^{9}$ cm$^{-3}$ 
    and Z = 1 Z$\subsun$.  The total cooling (solid) 
    includes all species contained within the Cloudy chemical network.  
    Components shown are CO (dot), 
    [Fe\textsc{I}] 24.0 $\mu$m (dash), 
    [Fe\textsc{II}] (long dash), 
    [O\textsc{I}] 63.2 $\mu$m (dash-dot), and 
    O\textsc{I} 630 nm.
  } \label{fig:cooln9_Z0}
\end{figure}

\begin{figure}
  \psfig{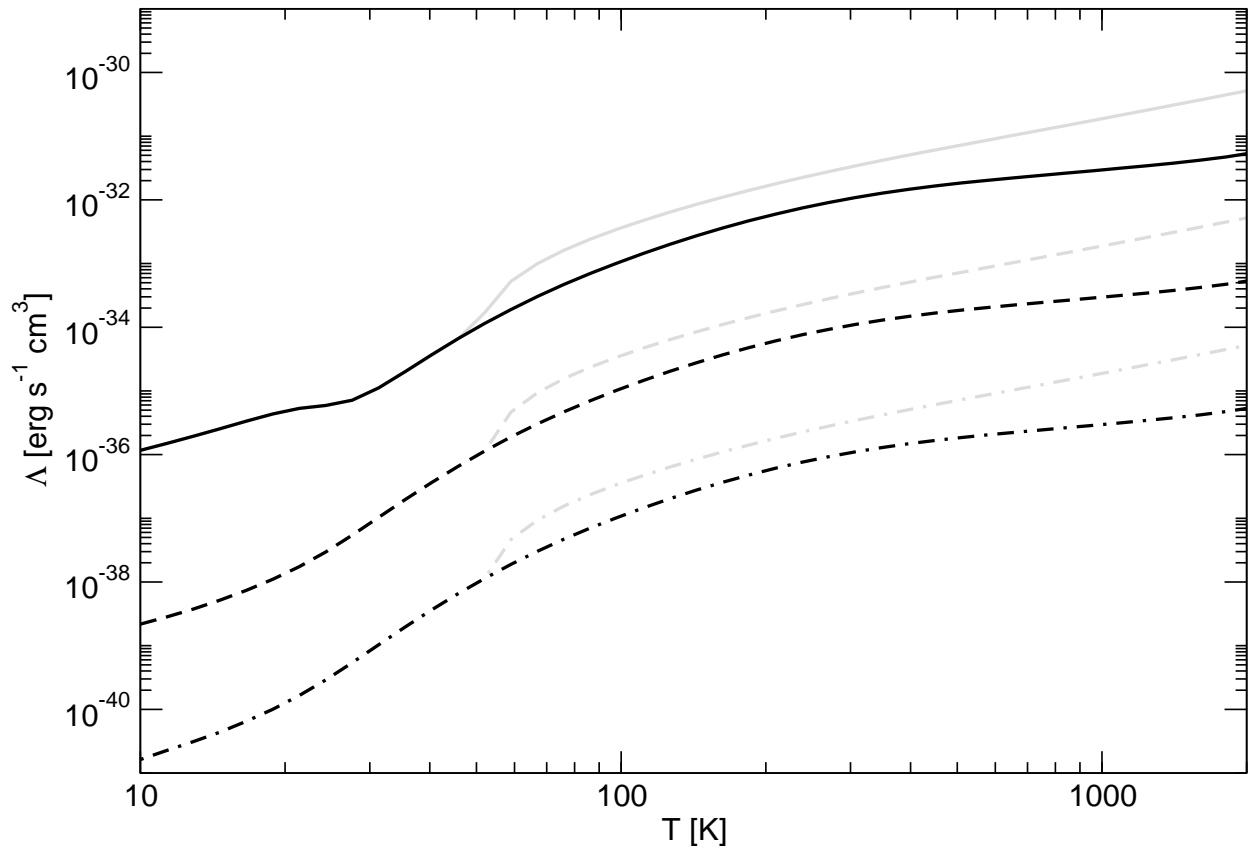}
  \caption{Total cooling rate from metals for gas at n$_{H}$ = 10$^{9}$ cm$^{-3}$, with 
    metallicities Z = 10$^{-2}$ Z$\subsun$ (solid), 10$^{-4}$ Z$\subsun$ (dashed), and 
    10$^{-6}$ Z$\subsun$ (dash-dot).  The black lines indicate the total cooling from 
    gas-phase metals only.  The grey lines show the cooling with gas-phase metals and 
    dust grains, created with the ISM dust grain model in Cloudy, using the command, 
    \texttt{grains ISM}.  The ISM dust model in Cloudy consists of 
    both graphite and silicates with sizes ranging from 5 $\times$ 10$^{-3}$ $\mu$m to 
    0.25 $\mu$m and a power-law size-distribution with a power-law index of -3.5.  
    For solar metallicity, the total grain abundances per H are 10$^{-9.811}$ for 
    graphite and 10$^{-9.748}$ for silicates.  In each case shown, the dust grain 
    abundances have been scaled to the gas-phase abundances.
  } \label{fig:cool_dust}
\end{figure}

In Figs. \ref{fig:COcooln3_Z-2}--\ref{fig:FeSSicooln3_Z0}, we plot 
the individual cooling contributions for number density, n$_{H}$ = 
10$^{3}$ cm$^{-3}$ and metallicities, Z = 10$^{-2}$ Z$\subsun$ 
and 1 Z$\subsun$.  
In each case, we create a set of cooling data 
with the full complement of elemental species, from H through Zn, 
neglecting H$_{2}$.  We plot only the 
coolants whose contributions reach, at least, 10$^{-3}$ of the total cooling 
within the temperature range, 10 K $\le$ T $\le$ 5000 K.  Since the cooling data 
were made assuming no incident ionizing radiation, as described in 
\S\ref{sec:cloudy}, we observe the dominant C transitions to be from C\textsc{I}, 
instead of C\textsc{II}, as in \citet{2003Natur.425..812B} and 
\citet{2006ApJ...643...26S}.  For Z = 10$^{-2}$ Z$\subsun$, the most important 
coolants are the fine structure transtions at 369.7 $\mu$m and 609.2 $\mu$m from 
C\textsc{I} and at 63.2 $\mu$m from O\textsc{I}.  At higher temperatures (T $\ge$ 
100 K), the Si\textsc{II} transition at 34.8 $\mu$m becomes important as well.  
At higher metallicities, CO replaces atomic C (Fig. \ref{fig:COcooln3_Z0}) 
and emission from Si\textsc{I} at 129.6 $\mu$m becomes dominant 
(Fig. \ref{fig:FeSSicooln3_Z0}).
Fe cooling is relatively unimportant up to n$_{H}$ = 
10$^{5}$ cm$^{-3}$, but is completely dominant for T $\ge$ 200 K by 
n$_{H}$ = 10$^{9}$ cm$^{-3}$, with O\textsc{I} strongest slightly below that 
temperature, and CO the most important below 50 K (Fig. \ref{fig:cooln9_Z0}).  
In addition to the elements covered by \citet{2006ApJ...643...26S}, we note 
that cooling from neutral S reaches roughly the 10\% level at 
n$_{H}$ = 10$^{3}$ cm$^{-3}$.  The [S\textsc{I}] 25.2 $\mu$m transition 
peaks at 40\% of the total cooling at n$_{H}$ = 10$^{7}$ cm$^{-3}$ and T $\sim$ 
1000 K.  The only other coolant that 
contributes at the level of at least 10$^{-3}$ of the total cooling is the 
60.6 $\mu$m transition of [P\textsc{II}] at n$_{H}$ = 10$^{5}$ cm$^{-3}$.  The 
number of coolants that reach 10$^{-3}$ of the total grows quickly with 
density.  We observe 23 distinct coolants contributing at that level at 
n$_{H}$ = 10$^{6}$ cm$^{-3}$, and 32 by 10$^{9}$ cm$^{-3}$.  
If we lower the threshhold to 10$^{-6}$, there are a total of 84 coolants at 
n$_{H}$ = 10$^{9}$ cm$^{-3}$, illustrating the strength of Cloudy and our 
cooling method.

\subsection{Dust Grains} \label{sec:dust}

Recently, studies have suggested that dust cooling at high densities can 
trigger fragmentation for metallicities as low as 10$^{-6}$ $Z\subsun$ 
\citep{2005ApJ...626..627O,2006MNRAS.369.1437S,2006ApJ...642L..61T,
2007arXiv0706.0613C}.  
\citet{2004MNRAS.351.1379S} have claimed that between 15 and 30\% of the mass 
of the progenitor of a pair-instability supernova is converted in dust.  However, 
observations of the Crab nebula \citep{2004MNRAS.355.1315G} and the Cassiopeia A 
supernova remnant \citep{2004Natur.432..596K} have returned little or no signs of 
dust, suggesting that type II supernova may not actually be large dust producers.  
Given the controversy surrounding the existence of dust grains in the 
formation environments of second-generation stars, we do not include the cooling 
from dust in the analysis, but leave it for a separate work.  To provide an 
example of the effect of dust on the cooling rate, we run a simple model including 
dust in Cloudy.  We use a model designed to simulate the dust within the ISM, using 
the Cloudy command, \texttt{grains ISM}.  The dust physics used in Cloudy is described 
in detail by \citet{2004MNRAS.350.1330V}.  The ISM dust model in Cloudy consists of 
both graphite and silicates with sizes ranging from 5 $\times$ 10$^{-3}$ $\mu$m to 
0.25 $\mu$m and a power-law size-distribution with a power-law index of -3.5.  
For solar metallicity, the total grain abundances per H are 10$^{-9.811}$ for 
graphite and 10$^{-9.748}$ for silicates.  In Fig. \ref{fig:cool_dust}, we plot 
the cooling rate from metals at n$_{H}$ = 10$^{9}$ cm$^{-3}$, with and without dust 
grains, for metallicities, Z = 10$^{-6}$ Z$\subsun$, 10$^{-4}$ Z$\subsun$, and 
10$^{-2}$ Z$\subsun$.  Since it is inappropriate to think of the dust and gas-phase 
metal abundances as independent, we directly scale the dust abundances with the 
gas-phase metal abundancs.  For number densities lower than 10$^{9}$ cm$^{-3}$, 
the additional cooling from dust is almost negligible.  At higher densities, dust 
becomes more important.  If dust is, in fact, produced in the supernovae of 
the first stars, it is likely to be as important as previous studies have claimed.

\subsection{Thermal Instability and Fragmentation} \label{sec:thermal}

\begin{figure}
  \hbox{
    \psfig{figure=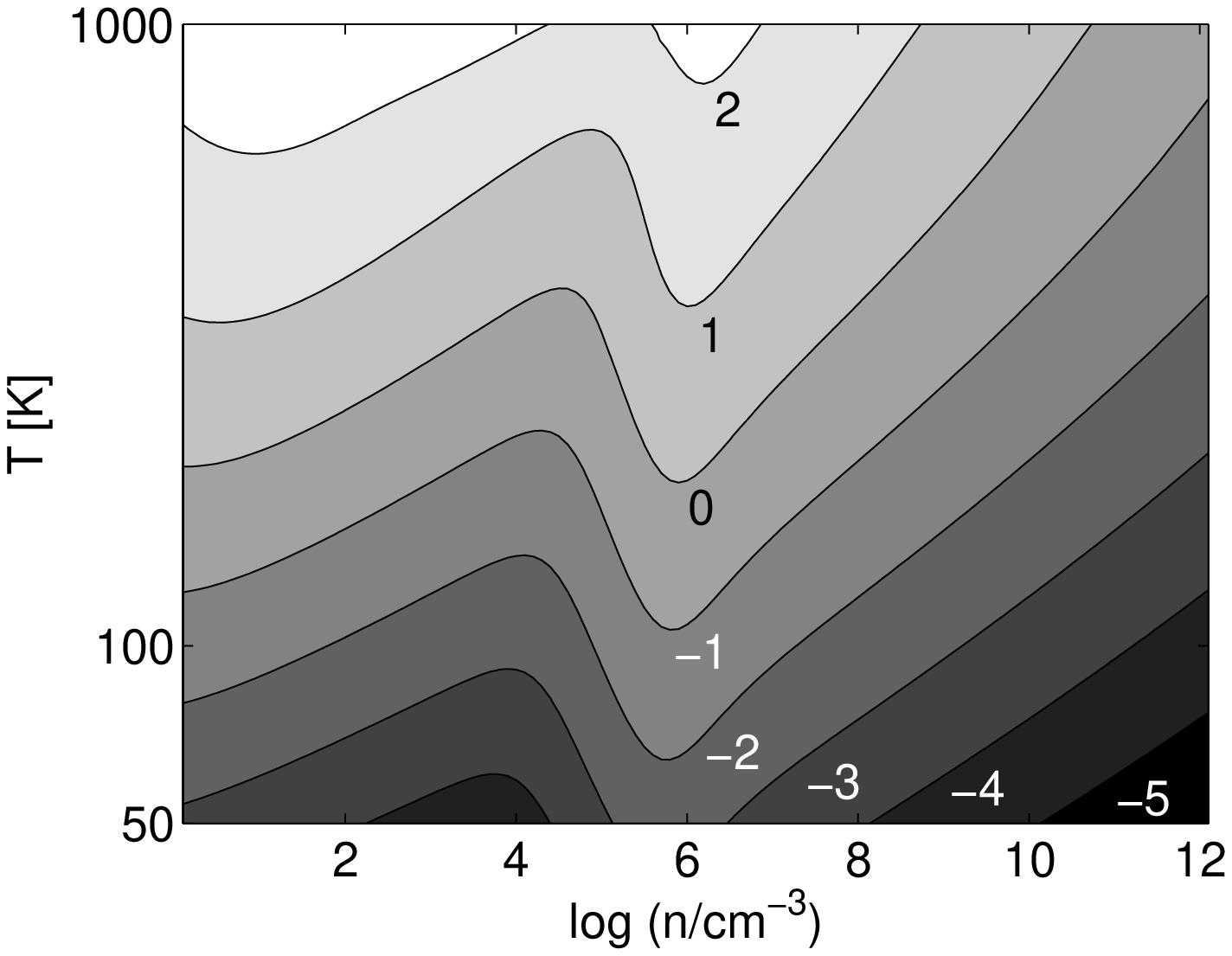,width=3.5in}
    \psfig{figure=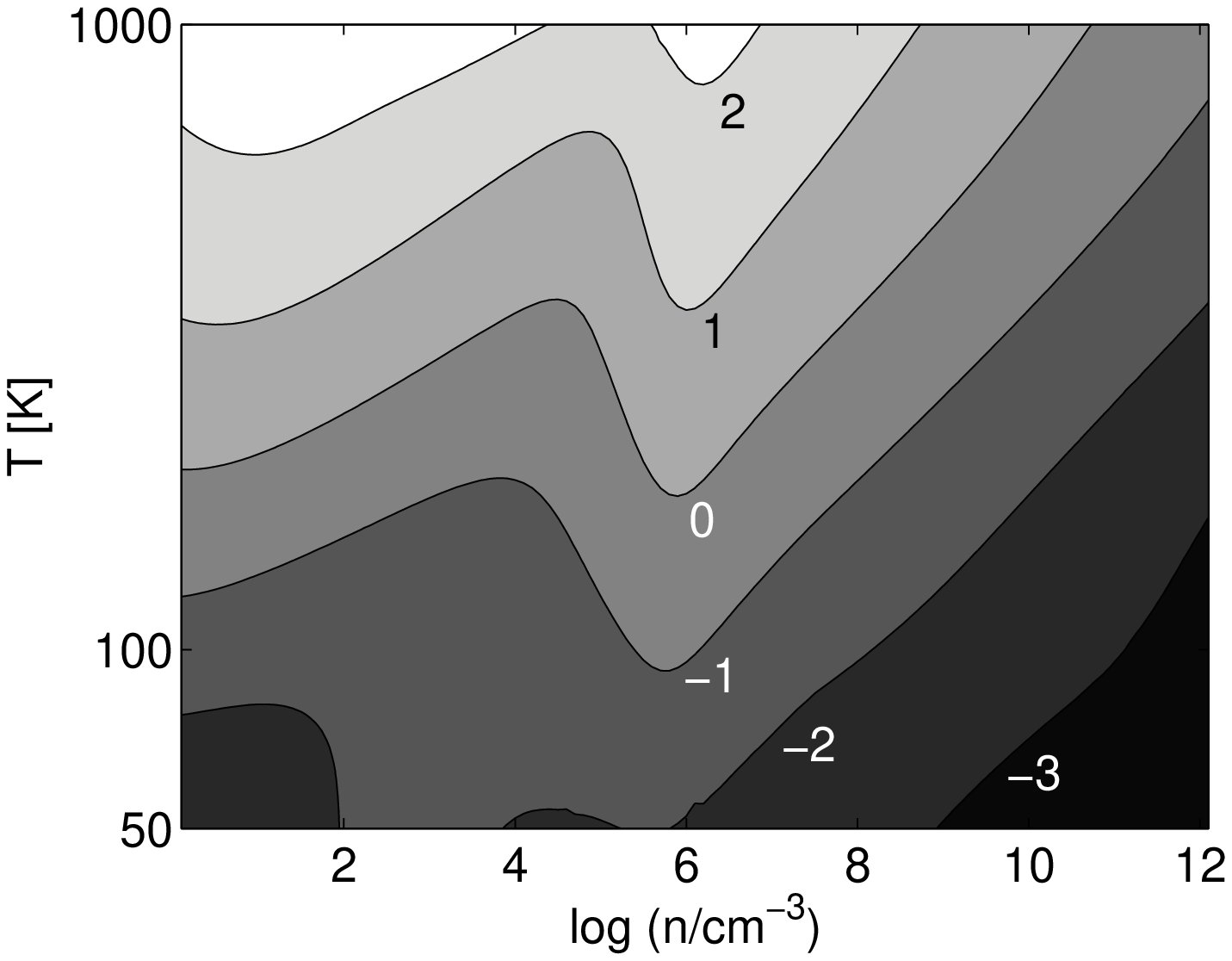,width=3.5in}
  }
  \hbox{
    \psfig{figure=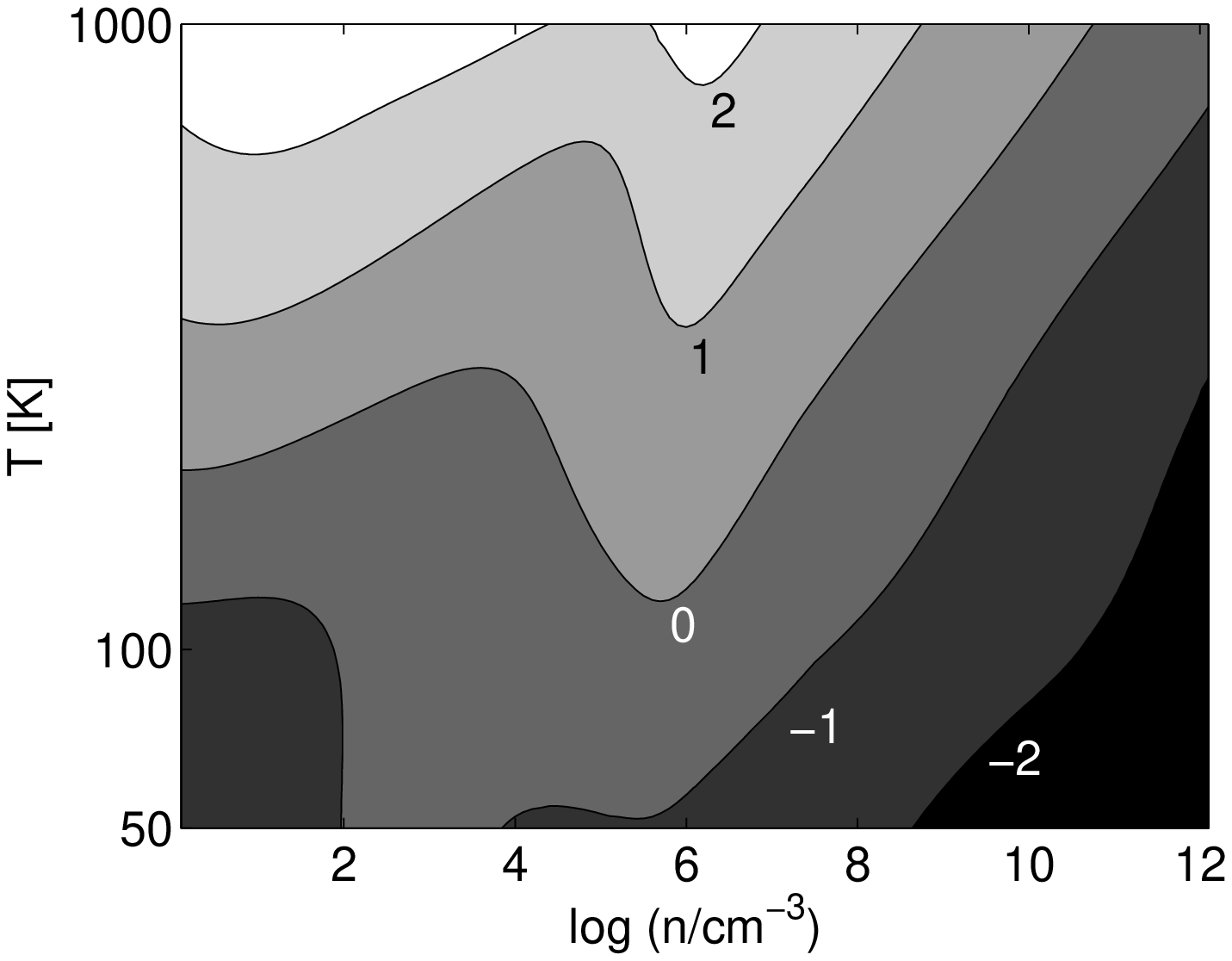,width=3.5in}
    \psfig{figure=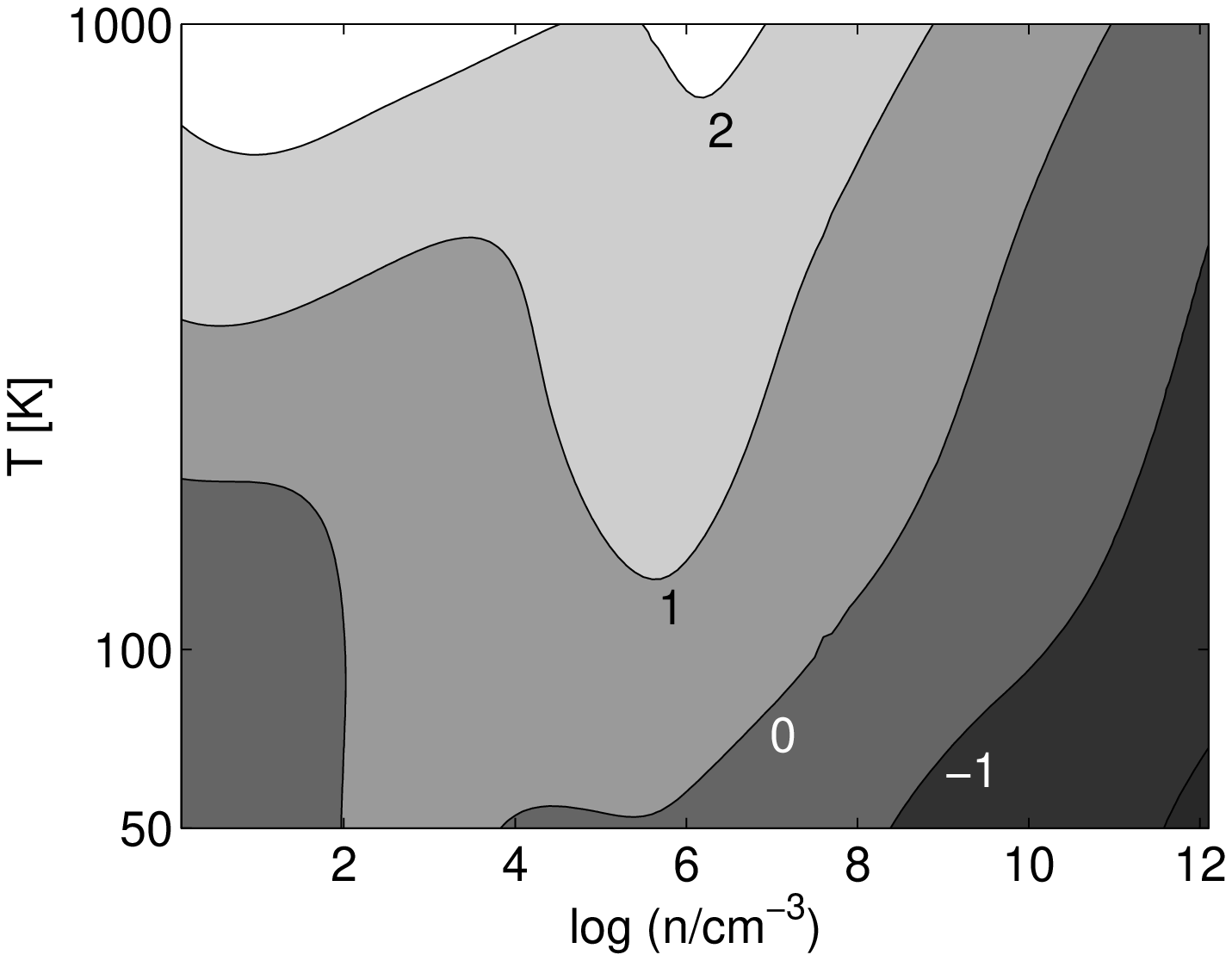,width=3.5in}
  }
  \hbox{
    \psfig{figure=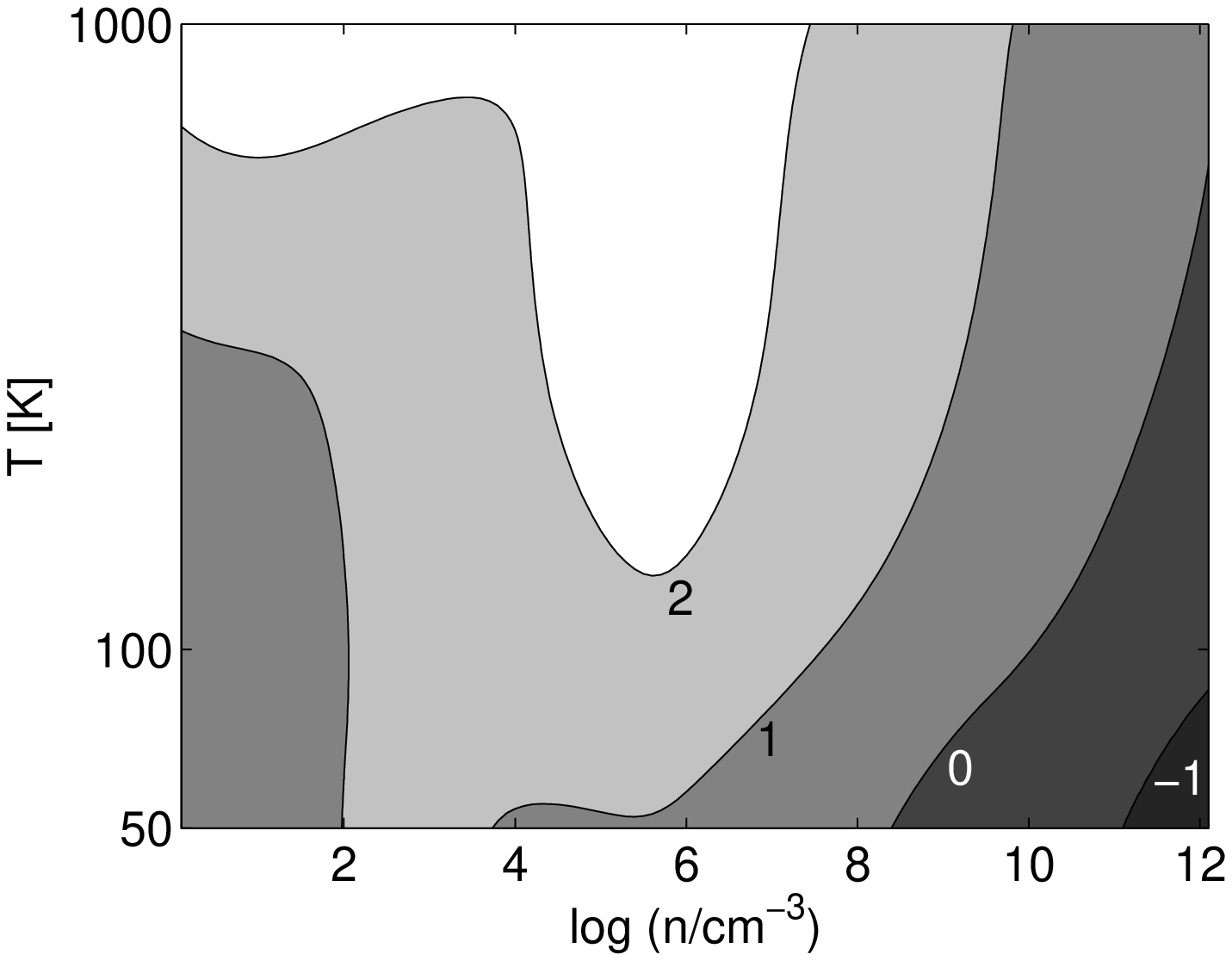,width=3.5in}
    \psfig{figure=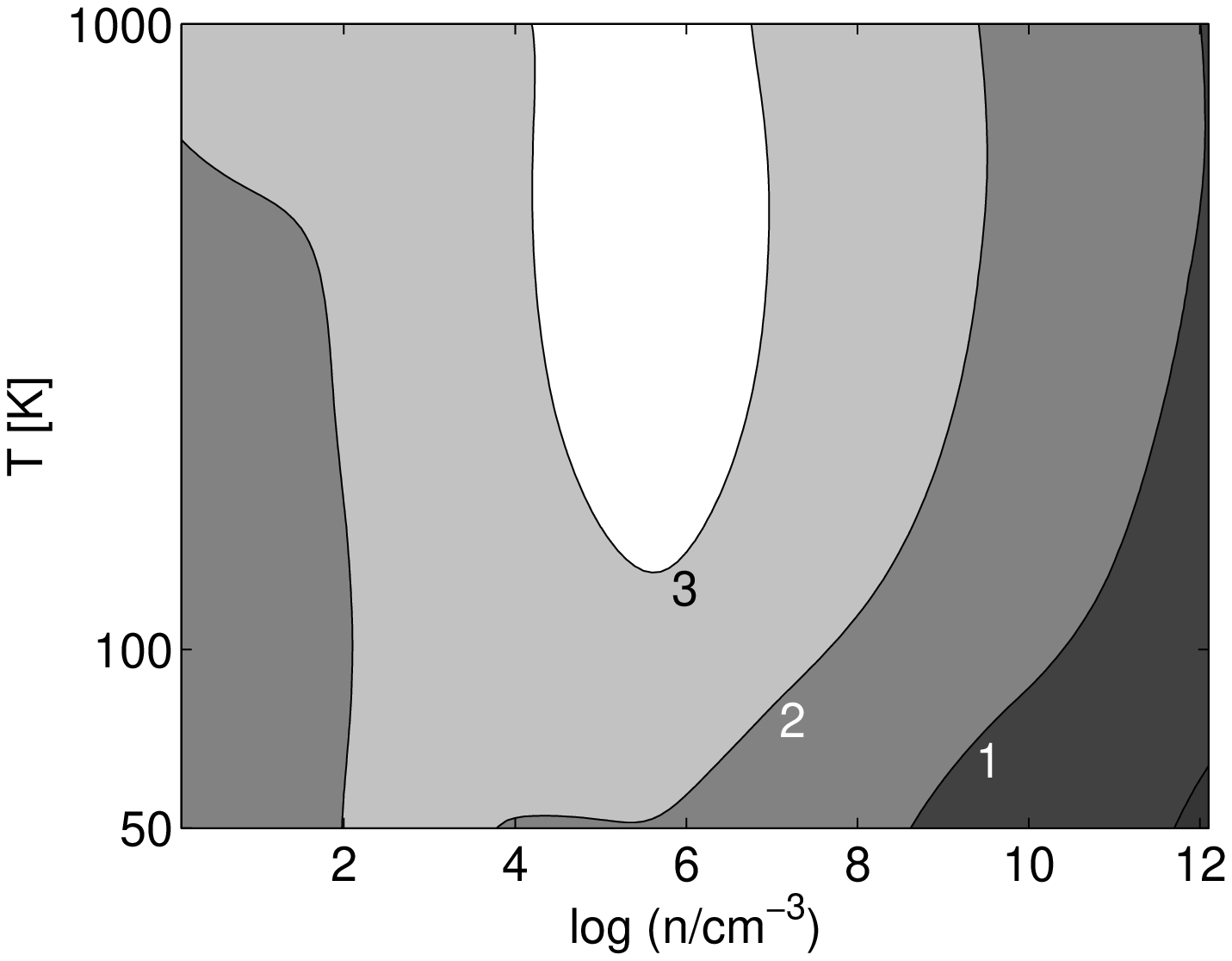,width=3.5in}
  }
  \caption{Contours of log$_{10}$(t$_{dyn}$/t$_{cool}$) over number density and 
    temperature 
    for gases with metallicities, Z = 0 (top-left), 10$^{-6}$ Z$\subsun$ 
    (top-right), 10$^{-5}$ Z$\subsun$ (middle-left), 10$^{-4}$ Z$\subsun$ 
    (middle-right), 10$^{-3}$ Z$\subsun$ (bottom-left), and 10$^{-2}$ Z$\subsun$ 
    (bottom-right).  H$_{2}$ cooling is included.
  } \label{fig:tdtc}
\end{figure}

In order to study the ability of a collapsing gas-cloud to fragment, we first 
identify the regions of density and temperature where the classical fragmentation 
criterion, t$_{cool}$ $<$ t$_{dyn}$ \citep{1965ApJ...142..531F}, is met, with the 
dynamical time expressed as 
\begin{equation} \label{eqn:tdyn}
t_{dyn} = \sqrt{\frac{3\pi}{16G\rho}}.
\end{equation}
We limit this analysis to solar abundance patterns.  
This represents the first step of an incremental approach to 
studying the general criteria that lead to fragmentation in collapsing 
clouds.  The use of solar abundance patterns will allow us to begin to quantify 
the chemical abundance required for fragmentation.  In a following paper, we 
will study nonsolar abundance patterns motivated by predicted yields of primordial 
supernovae.  In this future work, we will vary the abundances of individual elements, 
which will require the exploration of a much larger parameter-space.  
We create a 
Cloudy/ROCO data-grid with the following parameters: 50 K $\le$ T $\le$ 1000 K 
with $\delta$log(T) $\simeq$ 0.012 (100 points), 1 cm$^{-3}$ $\le$ n$_{H}$ $\le$ 
10$^{12}$ cm$^{-3}$ with $\delta$log(n$_{H}$) = 0.1, and 10$^{-6}$ Z$\subsun$ 
$\le$ Z $\le$ 10$^{-2}$ Z$\subsun$ with $\delta$log(Z) = 1.  We, then, follow the 
same procedure used to produce Fig. \ref{fig:cooling2}, as described in 
\S\ref{sec:enzo}.  In this section, we omit the decrease in the cooling rate caused 
by Compton heating on the CMB.  
In Fig. \ref{fig:tdtc}, we plot the log of the ratio of the 
dynamical time to the cooling time for each of the metallicities in the data-grid 
and for the metal-free case.  A cloud is able to fragment when  
log$_{10}$(t$_{dyn}$/t$_{cool}$) $>$ 0.  As expected, there is no density at which 
metal-free gas can fragment for T $<$ 200 K.  As the metallicity increases, the 
value of log$_{10}$(t$_{dyn}$/t$_{cool}$) slowly increases, first in the 
low-temperature regime, where H$_{2}$ cooling is inefficient, so even a small 
amount of metals has an effect.  For gas with n$_{H}$ = 10$^{4}$ cm$^{-3}$ at T = 
200 K, the fragmentation criterion is nearly met by Z = 10$^{-5}$ Z$\subsun$ and 
well satisfied one order of magnitude higher in metallicity.  Once the metallicity 
reaches 10$^{-2}$ Z$\subsun$, the entire parameter-space is fragmentable.  At 
high densities, however, fragmentation will be curtailed as the cloud becomes 
optically thick to its own radiation \citep{1976MNRAS.176..367L,1976MNRAS.176..483R}.  
As was also reported by \citet{2006ApJ...643...26S}, the efficiency of the metal 
cooling peaks at n$_{H}$ $\sim$ 10$^{6}$ cm$^{-3}$, significantly lowering the 
critical metallicity required for fragmentation.

\begin{figure}
  \hbox{
    \psfig{figure=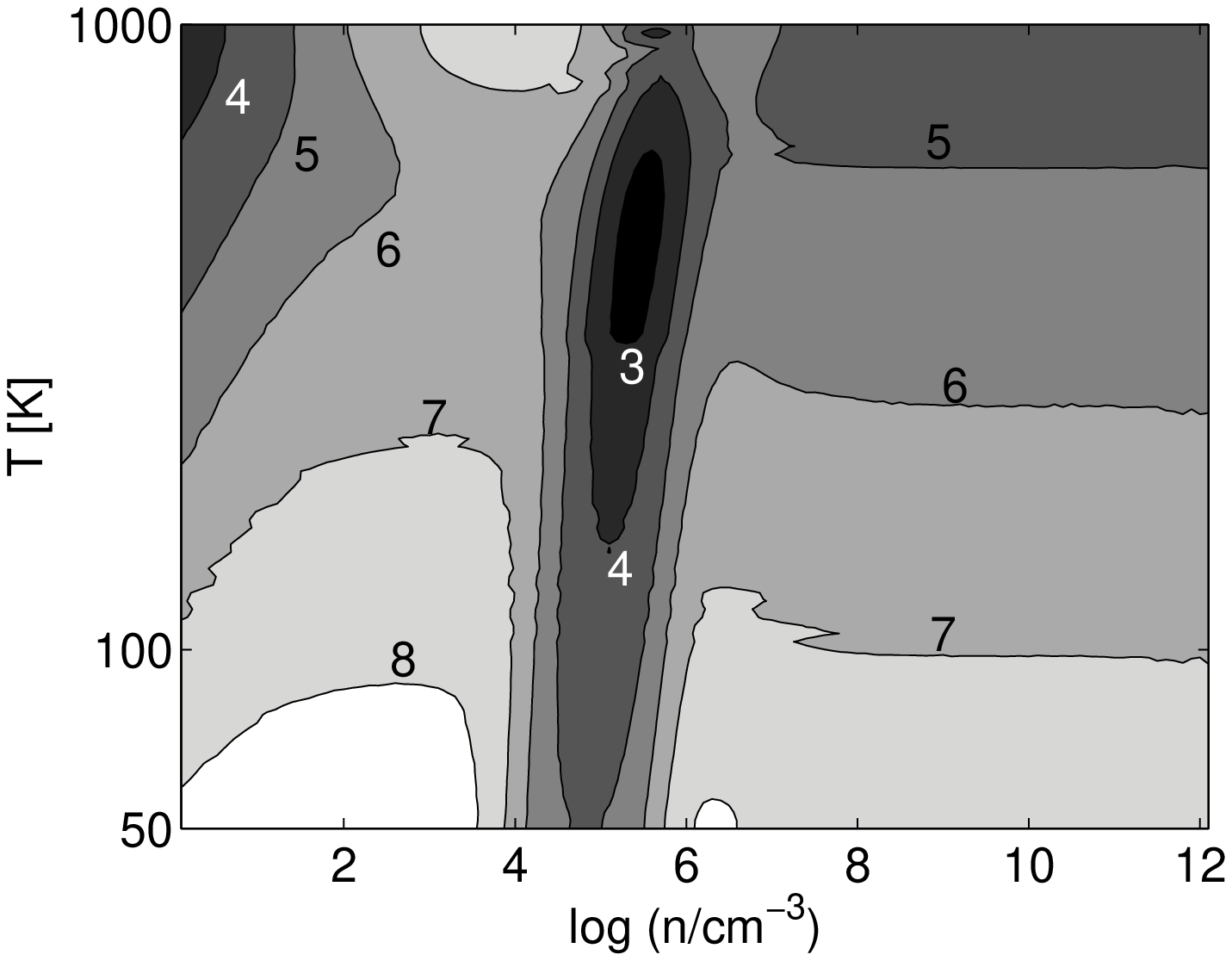,width=3.5in}
    \psfig{figure=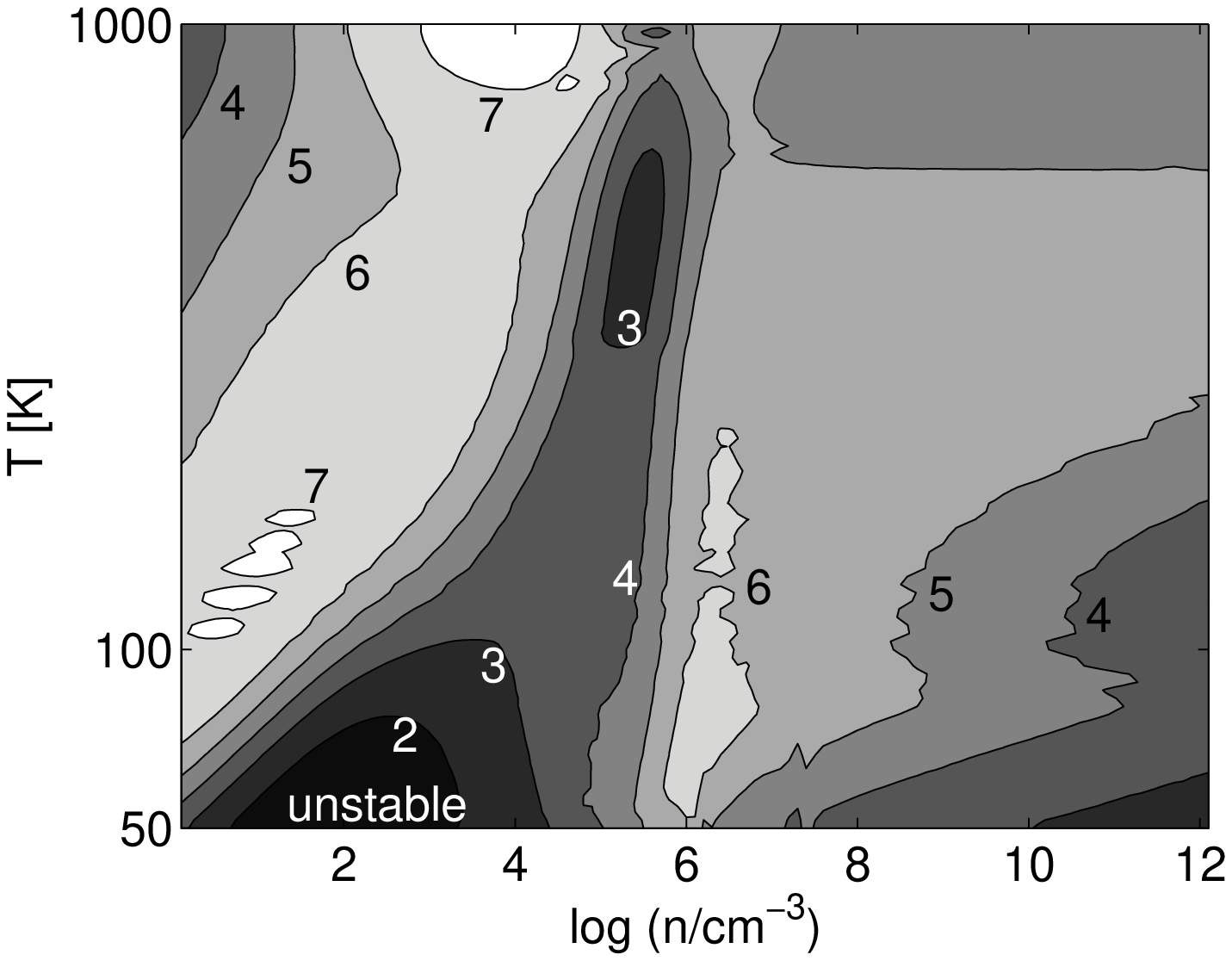,width=3.5in}
  }
  \hbox{
    \psfig{figure=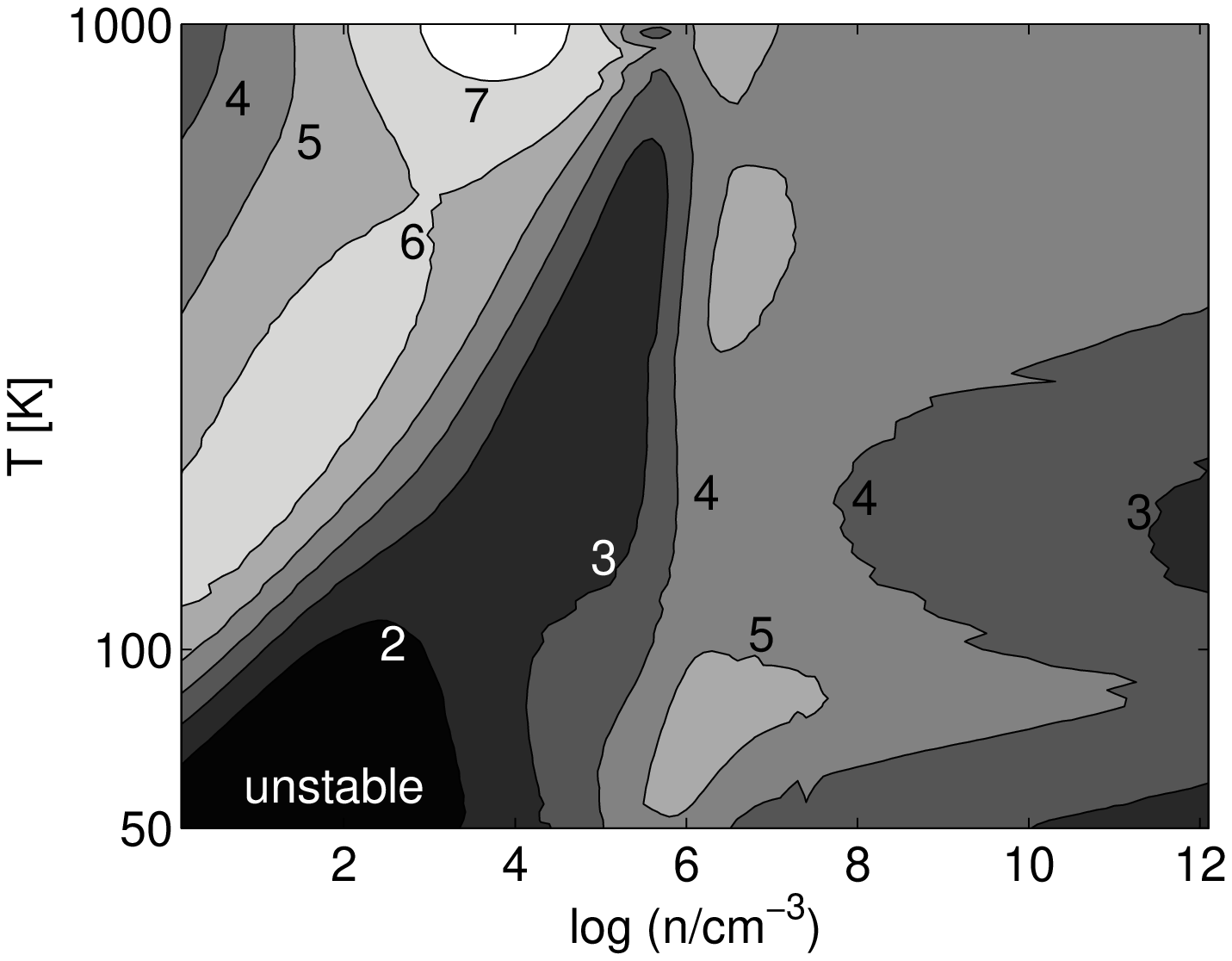,width=3.5in}
    \psfig{figure=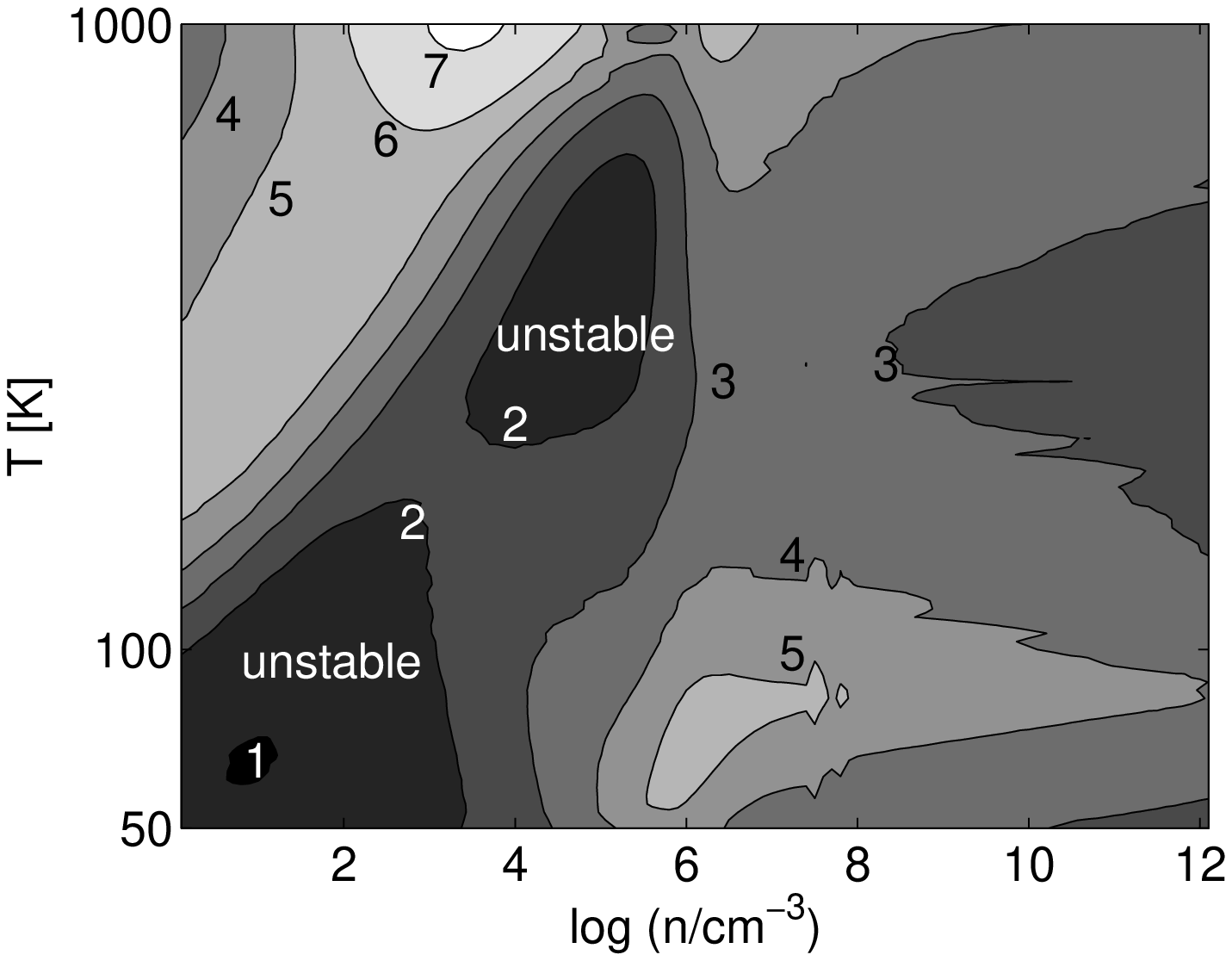,width=3.5in}
  }
  \hbox{
    \psfig{figure=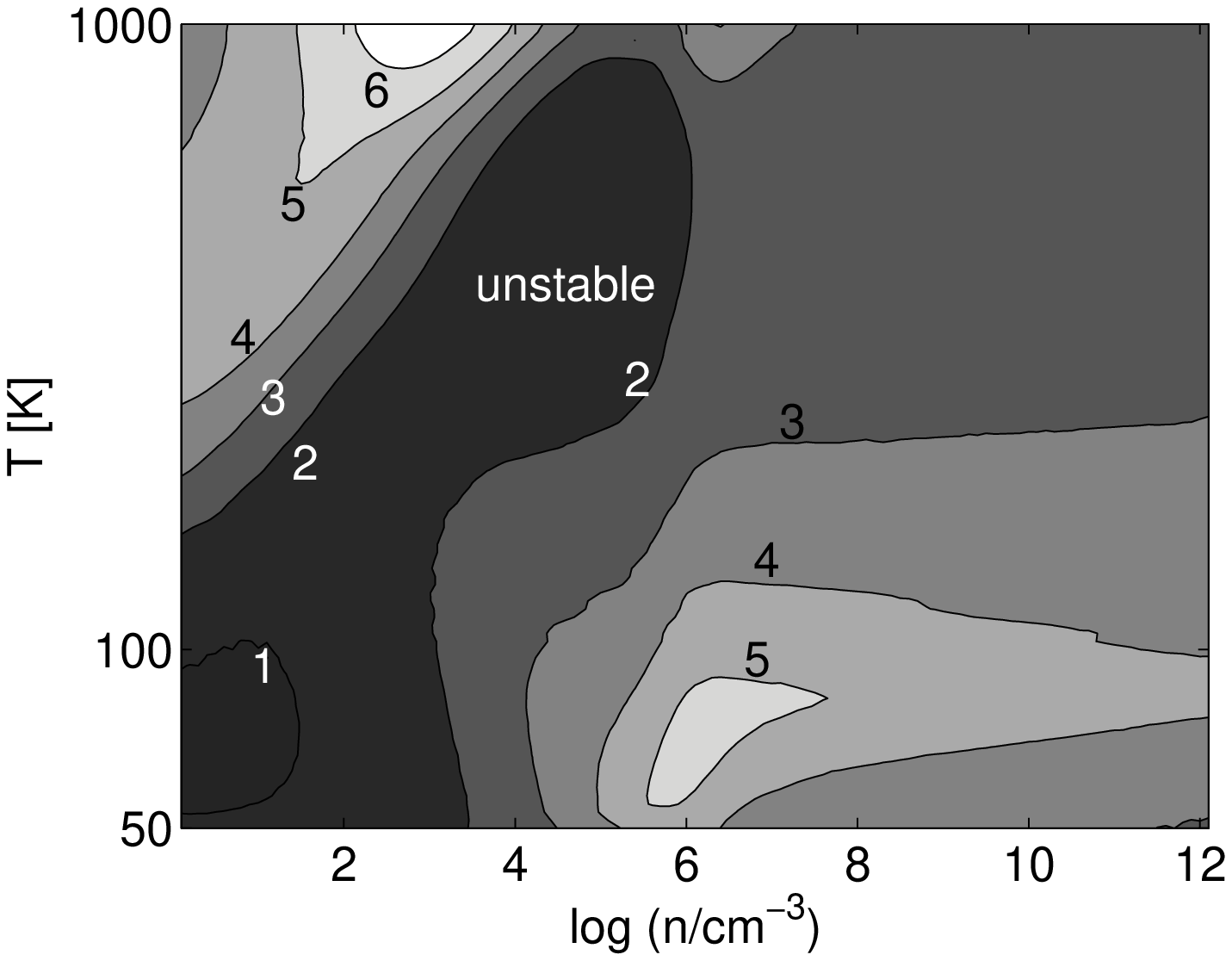,width=3.5in}
    \psfig{figure=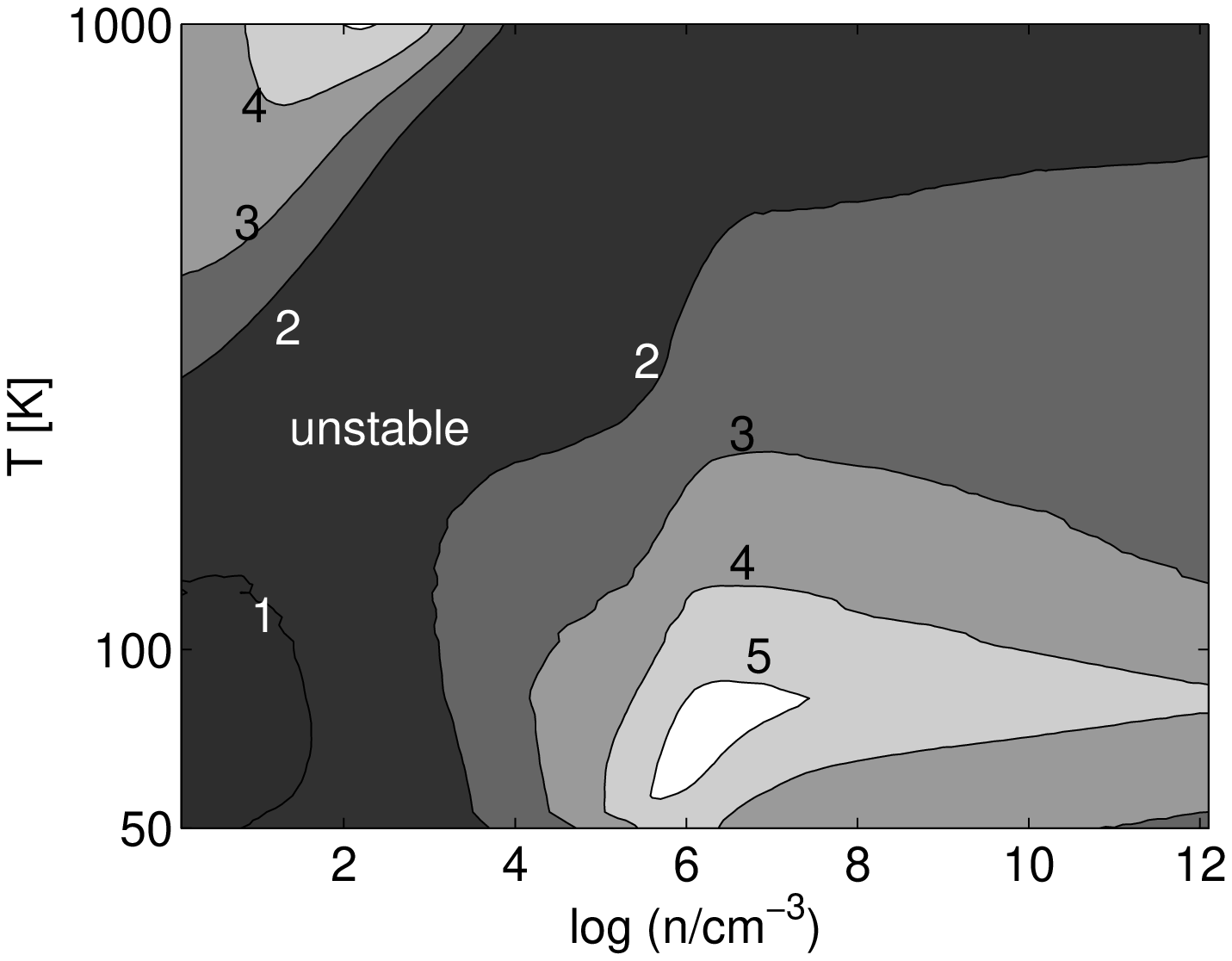,width=3.5in}
  }
  \caption{Contours of the instability parameter, ($\alpha$ - $\beta$) over 
    number density and temperature.  The medium is unstable for values less than 
    2.  The metallicities are Z = 0 (top-left), 10$^{-6}$ Z$\subsun$ 
    (top-right), 10$^{-5}$ Z$\subsun$ (middle-left), 10$^{-4}$ Z$\subsun$ 
    (middle-right), 10$^{-3}$ Z$\subsun$ (bottom-left), and 10$^{-2}$ Z$\subsun$ 
    (bottom-right).  At Z = 10$^{-4}$ Z$\subsun$, two separate thermally unstable 
    regions exist.  These two regions merge by 10$^{-3}$ Z$\subsun$.
  } \label{fig:ctStability}
\end{figure}

The addition of metals to a gas also has the potential to trigger thermal 
instabilities during cloud-collapse.  As in \citet{2002Sci...295...93A} 
\citep{1965ApJ...142..531F}, we 
define a parcel of gas losing energy at a rate, L, to be thermally unstable if 
\begin{equation} \label{eqn:thermal}
\rho \Big(\frac{\partial L}{\partial \rho}\Big)\Big|_{T} -
T \Big(\frac{\partial L}{\partial T}\Big)\Big|_{\rho} + L(\rho,T) > 0,
\end{equation}
where L is expressed in terms of the cooling rate, $\Lambda$, as
\begin{equation} \label{eqn:L}
L(\rho,T) = \rho \Lambda(\rho,T).
\end{equation}
The cooling rate, $\Lambda$, can be locally approximated by a power-law in both 
temperature and density as
\begin{equation} \label{eqn:powerLaw}
\Lambda(\rho,T) \propto \Big(\frac{T}{T_{0}}\Big)^{\alpha} 
\Big(\frac{\rho}{\rho_{0}}\Big)^{\beta}.
\end{equation}
The partial derivatives of Eqn. \ref{eqn:thermal} become 
\begin{equation} \label{eqn:thermalFirst}
\Big(\frac{\partial L}{\partial \rho}\Big)\Big|_{T} = (\beta + 1)\Lambda(\rho,T)
\end{equation}
and
\begin{equation} \label{eqn:thermalSecond}
\Big(\frac{\partial L}{\partial T}\Big)\Big|_{\rho} = 
\frac{\rho \alpha \Lambda(\rho,T)}{T}.
\end{equation}
The thermal instability criterion simplifies to
\begin{equation} \label{eqn:thermal2}
\alpha - \beta < 2.
\end{equation}
In Fig. \ref{fig:ctStability} we plot the value of the instability parameter, 
($\alpha$ - $\beta$), for the same cooling data used for Fig. \ref{fig:tdtc}.  
For metal-free gas (Fig. \ref{fig:ctStability}, top-left), the instability 
parameter is greater than 3 over nearly the entire parameter space, and always 
greater than 4 at high densities.  \citet{2002Sci...295...93A} and 
\citet{2006ApJ...652....6Y} both arrive at the same conclusion using this 
analysis for metal-free gas, with the added assumption that the cooling function 
is independent of density.  As the metallicity increases, a region of thermal 
instability forms at low density and temperature.  When the metallicity reaches 
10$^{-4}$ Z$\subsun$ (Fig. \ref{fig:ctStability}, middle-right), a second unstable 
region exists for 10$^{3}$ cm$^{-3}$ $\lesssim$ n$_{H}$ $\lesssim$ 10$^{6}$ 
cm$^{-3}$, at a temperature of a few hundred K.  The second unstable region 
coincides with the increase in cooling efficiency to its maximum value, illustrated 
in Fig. \ref{fig:tdtc}.

\begin{figure}
  \hbox{
    \psfig{figure=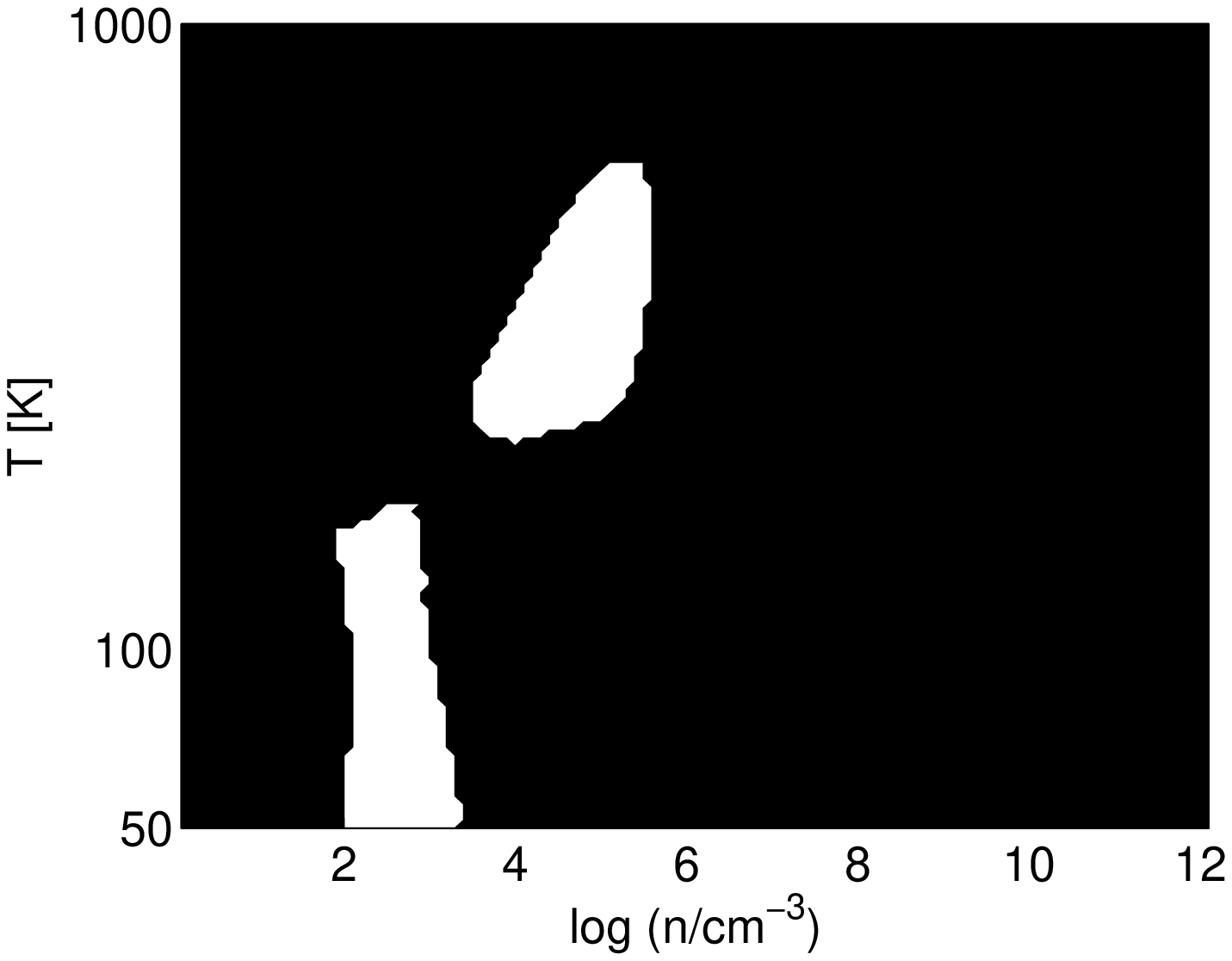,width=3.5in}
  }
  \hbox{
    \psfig{figure=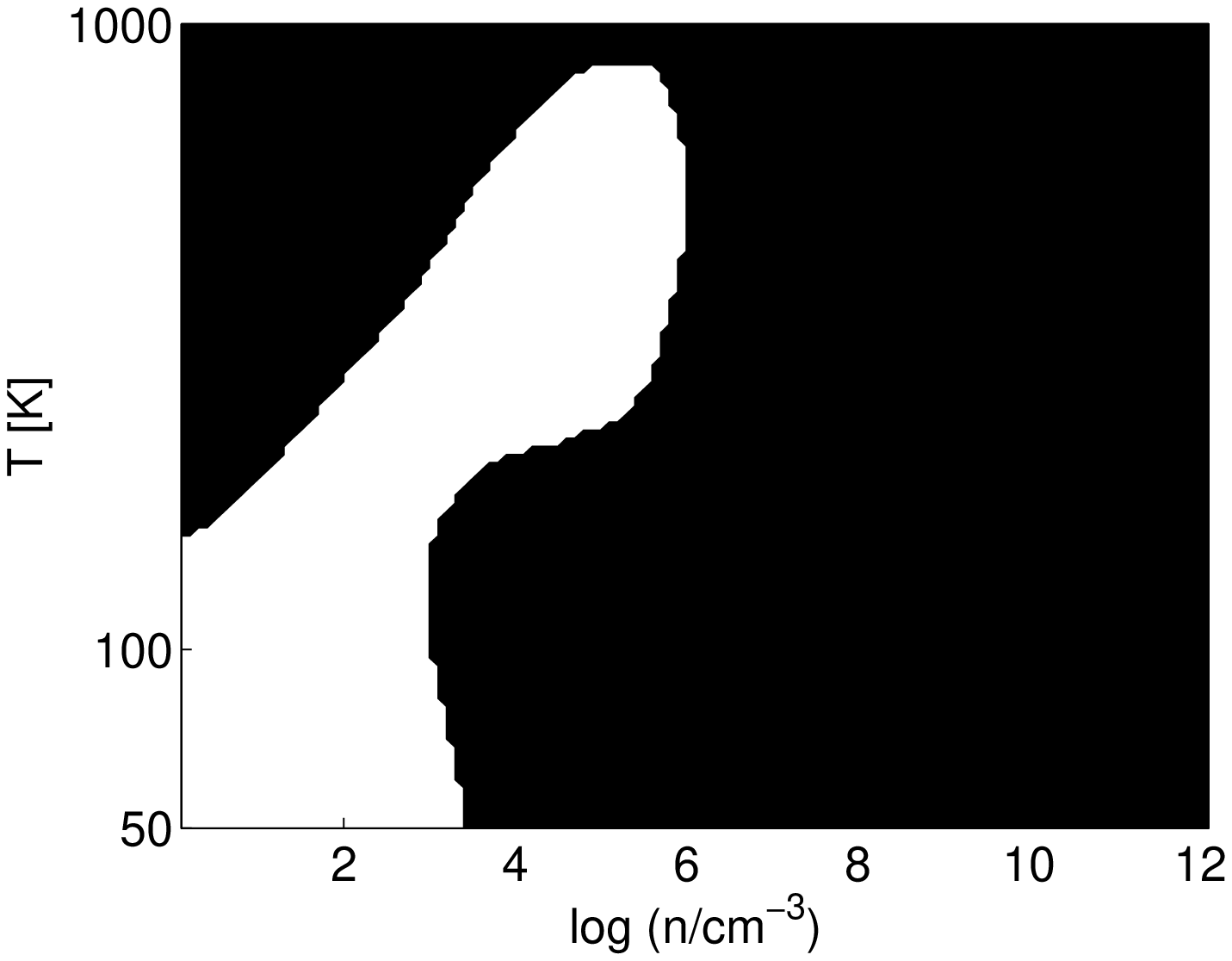,width=3.5in}
  }
  \hbox{
    \psfig{figure=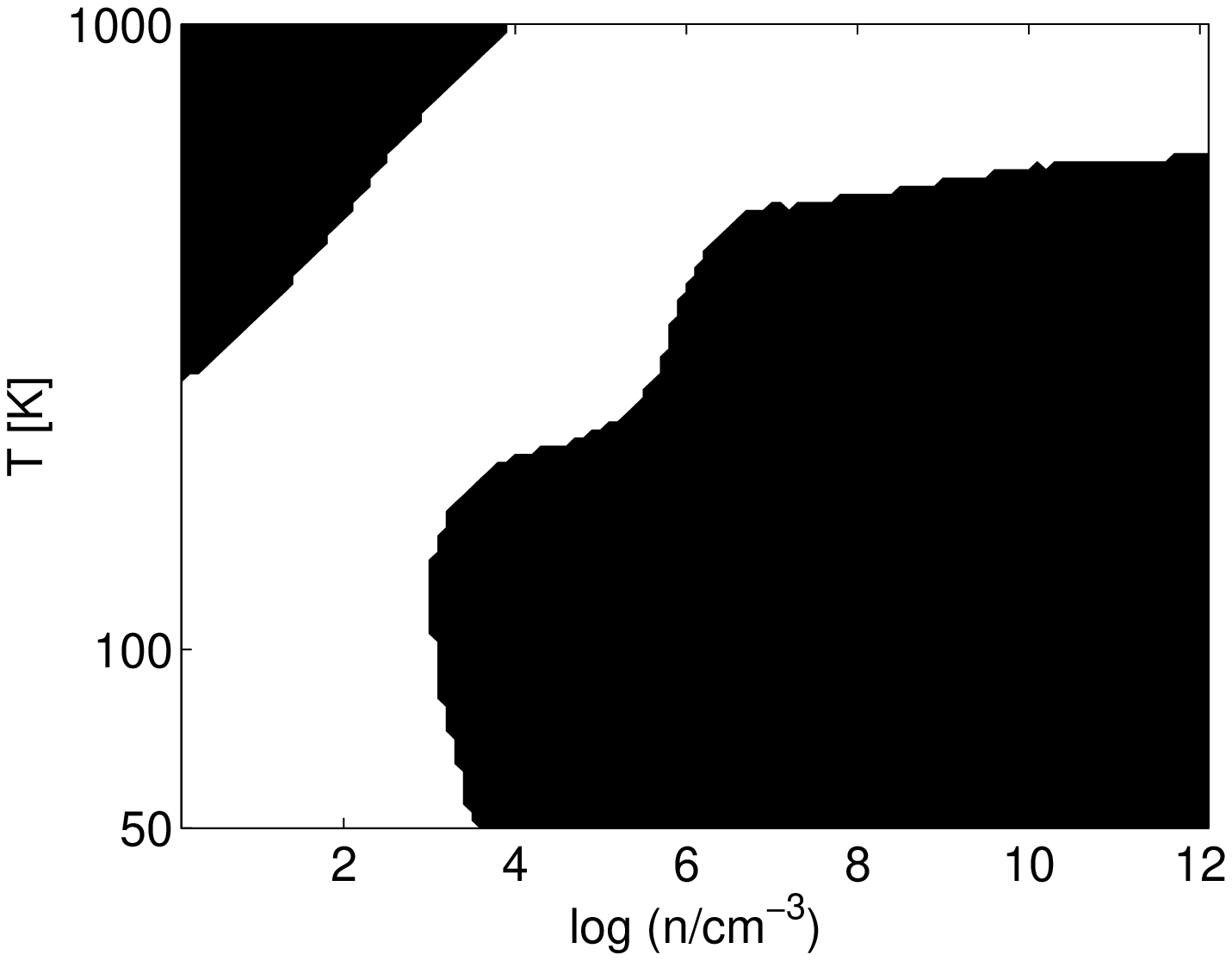,width=3.5in}
  }
  \caption{The white patches indicate regimes of density and temperature where 
  log$_{10}$(t$_{dyn}$/t$_{cool}$) $>$ 0 and ($\alpha$ - $\beta$) $<$ 2 for 
  metallicities, Z = 10$^{-4}$ Z$\subsun$ (top), 10$^{-3}$ Z$\subsun$ 
  (middle), and 10$^{-2}$ Z$\subsun$ (bottom).  As in Fig. 
  \ref{fig:ctStability}, there are two individual doubly unstable regions at 
  Z = 10$^{-4}$ Z$\subsun$ that have merged by 10$^{-3}$ Z$\subsun$.
  } \label{fig:doubleStability}
\end{figure}

Fragmentation is more likely to occur when the gas is both thermally 
unstable, and can cool faster than the dynamical time.  We indicate the regions 
where both the fragmentation and thermal instability criteria are met in white in 
Fig. \ref{fig:doubleStability}.  No doubly unstable realm exists for 
metallicities, Z $\le$ 10$^{-5}$ Z$\subsun$.

\subsection{Effects of the Cosmic Microwave Background} \label{sec:cmb}

\begin{figure}
  \hbox{
    \psfig{figure=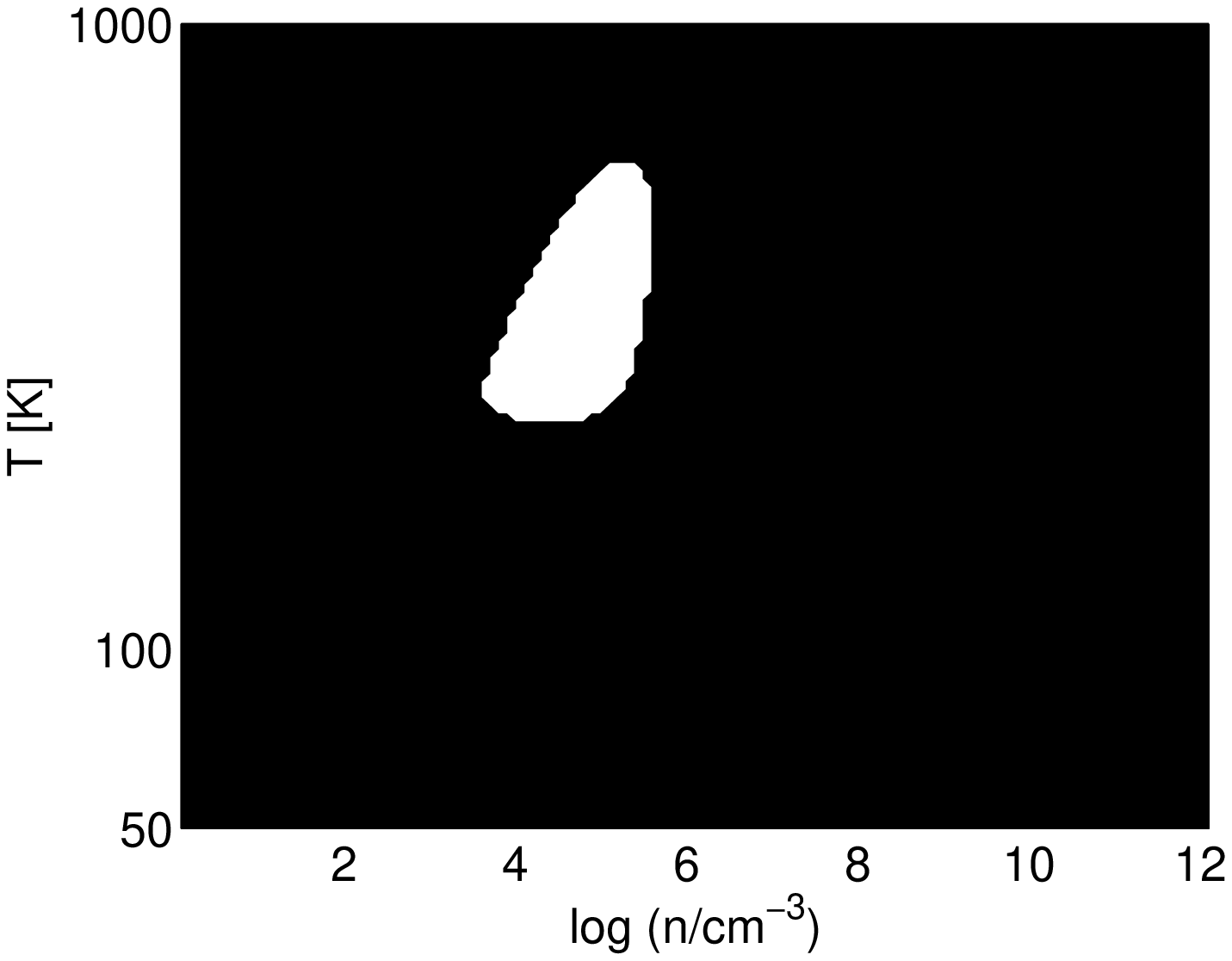,width=3.5in}
  }
  \hbox{
    \psfig{figure=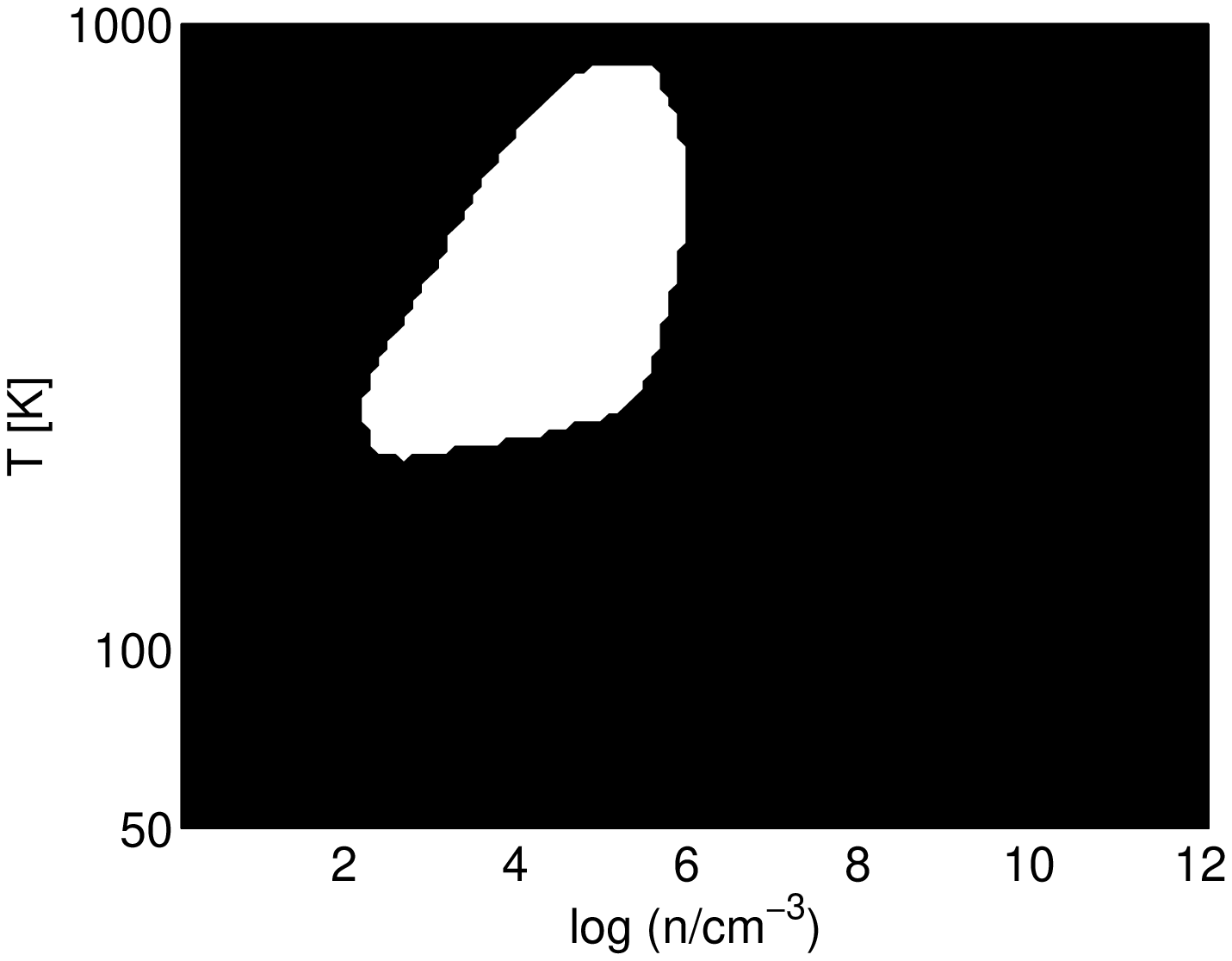,width=3.5in}
  }
  \hbox{
    \psfig{figure=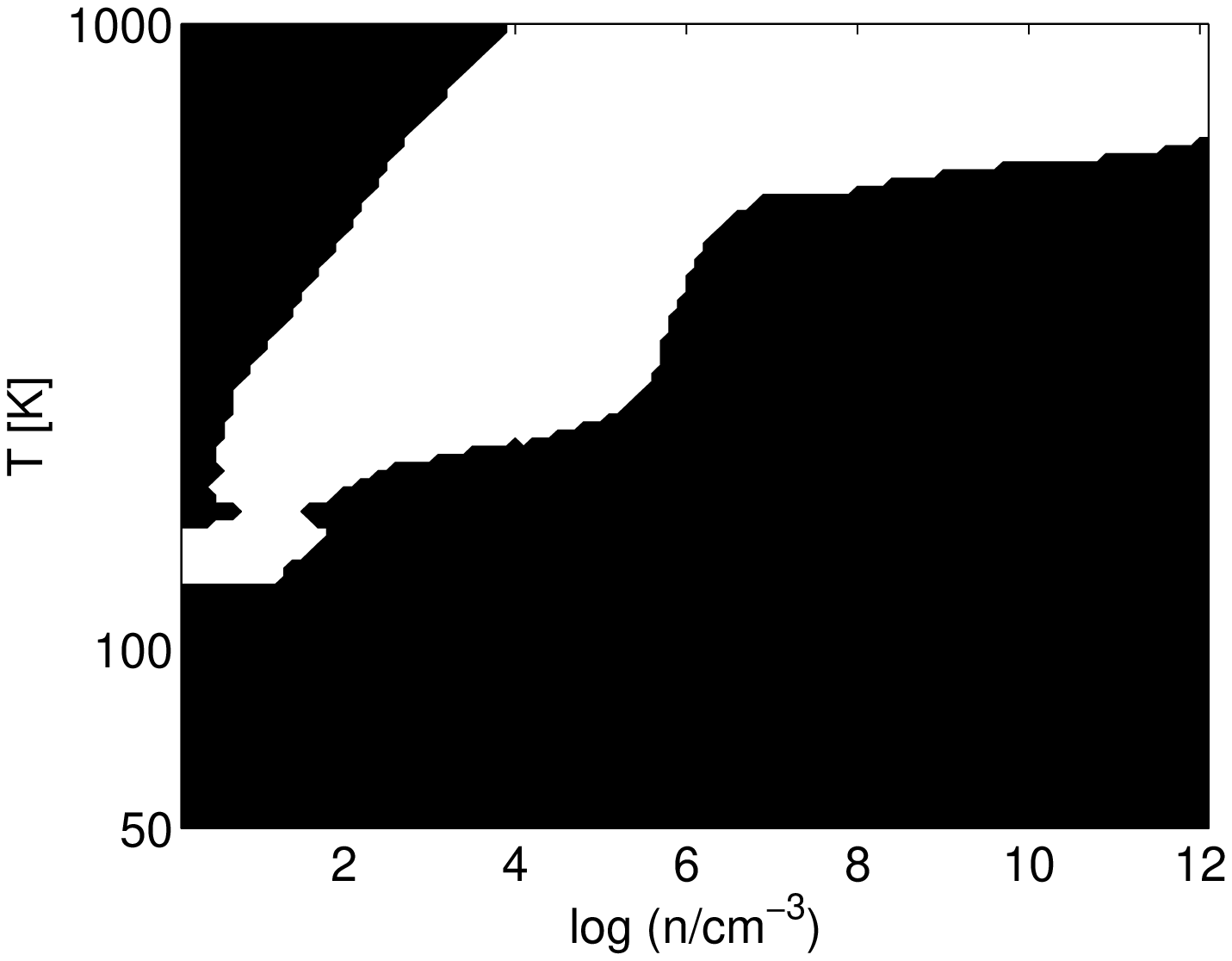,width=3.5in}
  }
  \caption{Doubly unstable regions for the same metallicities as in Fig. 
    \ref{fig:doubleStability}, but with a CMB temperature floor at z = 20 
    included.
  } \label{fig:doubleStabilityCMB}
\end{figure}

The CMB creates a temperature floor, below which gas cannot cool radiatively.  
We study the influence of the CMB on the evolution of star-forming gas by applying 
a CMB floor at z = 20 in the manner described in \S\ref{sec:enzo} to the cooling 
data-grid used in \S\ref{sec:thermal}.  The CMB affects the cooling properties of 
the gas in two ways.  The first is by increasing the cooling time at temperatures 
near the CMB temperature to greater than the dynamical time so that the 
fragmentation criterion is no longer satisfied.  The second is by increasing the 
value of $\alpha$, from Equation \ref{eqn:thermal2}, at low temperatures, 
making the gas thermally stable.  In Fig. \ref{fig:doubleStabilityCMB}, we 
illustrate the influence of the CMB on the doubly unstable regions, shown 
previously.  The unstable region that existed in the low-density, low-temperature 
regime is completely eliminated.  There remains, however, a small area of 
instability for metallicities as low as 10$^{-4}$ Z$\subsun$.

\section{Discussion} \label{sec:discussion}

We have introduced a new method for including the radiative cooling from metals 
in large, three-dimensional hydrodynamic simulations.  In addition to its 
implementation in the AMR code, Enzo, this method has also been used by 
\citet{2006AIPC..873..257B} in numerical simulations with the 
smoothed particle hydrodynamics (SPH) code, Gadget \citep{2001NewA....6...79S,
2005MNRAS.364.1105S}.  
Our technique takes 
advantage of the extremely complex chemical reaction network of the preexisting 
radiative transfer code, Cloudy, which includes a full elemental coverage from H to 
Zn, along with a variety of molecular species and dust.  With the singular 
assumption of ionization equilibrium for the heavy elements, we are able to 
precalculate cooling rates for gases with any chemical abundance in all manners 
of radiation environments over a temperature range of 10 to 10$^{8}$ K.  With 
cooling rates computed in advance, we eliminate the barrier that has classically 
prevented large chemical models from being incorporated into three-dimensional 
numerical simulations.  Because our cooling scheme is valid over such a large 
range of density and temperature, and features so many coolants, it can 
be applied to a huge variety of astrophysical problems, such as the evolution of 
the ISM and IGM, normal star formation, planetary nebulae, accretion disks, and 
protoplanetary disks.

One advantage of the large chemical network of Cloudy is that we are able to 
determine the dominant coolants from a complete sample of atomic species up to 
an atomic number of 30.  Fine-structure transitions of C and O are the greatest 
contributors to the cooling up to number densities of about 10$^{6}$ cm$^{-3}$, 
where Fe cooling becomes significant and C is marginalized.  The importance of 
these three elements, along with Si, in triggering the formation of the first 
low-mass stars has been studied in great detail by \citet{2006ApJ...643...26S}.  
The cooling models we have constructed using solar abundance patterns reveal S 
cooling to be important for 10$^{3}$ cm$^{-3}$ $\lesssim$ n$_{H}$ $\lesssim$ 
10$^{9}$ cm$^{-3}$.  S is produced only slightly less than Si in type II 
\citep{1995ApJS..101..181W} and pair-instability supernovae 
\citep{2002ApJ...567..532H}, and should be taken into account when considering 
the metals responsible for the transition from Population III to Population II 
star formation.  The ability to specify individual abundances in our cooling 
method makes it straightforward to simulate the evolution of gas with non-solar 
abundance patterns.

In addition to elevating the cooling rate of a gas to satisfy the classical 
fragmentation criterion, metals also increase the potential for fragmentation by 
creating thermal instabilities.  We have identified regions in temperature and 
density in which both the classical fragmentation and thermal instability criteria 
are met to be the physical conditions most likely to see fragmentation occur.  We 
observe these doubly unstable regions to exist for metallicities as low as 
10$^{-4}$ Z$\subsun$.  If fragmentation cannot occur outside these regions, then 
the fate of a star-forming gas-cloud will be determined by the path taken 
through density-temperature space as it collapses.  
If we consider the doubly 
unstable regions in Fig. \ref{fig:doubleStability}, appropriate for 
star-formation in current epoch, there will almost certainly be a period of 
double instability when Z $\ge$ 10$^{-3}$ Z$\subsun$.  Interestingly, the 
density-temperature tracks shown in Fig. 1 of \citet{2005ApJ...626..627O} 
indicate that gas with 
Z = 10$^{-4}$ Z$\subsun$ will pass right through the stable region that separates 
two instabilities.  \citet{2005ApJ...626..627O} also find that star formation 
at that metallicity only produces high-mass fragments.  At high redshift, the CMB 
significantly reduces the size of the doubly unstable regions.  Thermal 
instabilities, though, are extremely sensitive to the slope of the cooling rate 
as a function of density and temperature.  Every element has distinct cooling 
properties, and will, therefore, produce different thermal instabilities.  
As such, the key to uncovering the nature of the first Population II stars will be 
in the determination of the mass function of their precursors.

\section{Future Development} \label{sec:future}

In papers to follow, we will extend our study to gases with non-solar abundance 
patterns.  We will explore thermal and double instabilities created by individual 
elements, as well as abundance patterns produced in Population III supernovae.  
Future studies will also investigate the effects of background radiation on the 
evolution of star-forming gas.  
The final word, however, will only come from three-dimensional, numerical 
simulations.  The simulations by \citet{2007ApJ...661L...5S}, employing the 
methods described here with solar abundance patterns, have confirmed that 
fragmentation occurs for metallicities, Z $\ge$ 10$^{-3}$ Z$\subsun$.  
We intend to pair all future predictions made from analysis of thermal 
instabilities with full numerical simulations.  

Although dust physics has been implemented in Cloudy, we currently treat only 
metals in the gas-phase in our analysis.  In the future, we will study the effects 
of dust cooling in more detail.  We will also couple the dust chemistry to the H/He 
chemical network in Enzo, so as to properly model the formation of H$_{2}$ on 
grain surfaces.  
A great strength of this method is the use of the ever-expanding 
chemical network of Cloudy.  As more physical processes are incorporated into the 
Cloudy software, the utility of this method will increase as well.  
One major constraint of this work, however, is that its validity is confined to 
the optically thin limit.  The approximations made here break down at opacities 
of order unity.  For higher opacities, our method provides a core module for 
flux-limited diffusion schemes.  Complex geometries 
introduce problems of self-heating and shadowing, which will require full, 
three-dimensional radiative transfer.

\section*{Acknowledgments}
Based on observations made with the NASA/ESA Hubble Space Telescope,
obtained from the data archive at the Space Telescope Institute. STScI is
operated by the association of Universities for Research in Astronomy, Inc.
under the NASA contract NAS 5-26555.  This research was supported by 
Hubble Space Telescope Theory Grant HST-AR-10978.01, a grant from 
NASA's ATP NNG04GU99G, and NSF CAREER award AST-0239709 from the National 
Science Foundation.  We are grateful to an anonymous referee for 
sharing many insightful comments and suggestions.  
We also thank Brian O'Shea and Mike Norman for useful 
discussions and support.  SS also thanks KIPAC and Stanford University
for their hospitality.


\bsp

\clearpage

\clearpage

\clearpage

\clearpage

\clearpage

\clearpage

\clearpage

\clearpage

\clearpage

\clearpage

\clearpage

\clearpage

\label{lastpage}

\end{document}